\documentclass[twocolumn]{aastex63}

\usepackage{rotating}
\usepackage{acronym}
\usepackage{xspace}
\usepackage{multirow}
\usepackage{mathtools}
\usepackage{amsmath}
\usepackage{hyperref}
\usepackage{units}
\usepackage{tabularx}
\usepackage[normalem]{ulem}
\usepackage{xcolor}

\newcommand{\Teff}{\ifmmode {T_{\rm eff}}\else${T_{\rm eff}}$\fi}
\newcommand{\Msun}{\ensuremath{M_\odot}}

\newcommand{\posydon}{\texttt{POSYDON}\xspace}

\newcommand{\mesa}{\texttt{MESA}\xspace}

\newcommand{\Hecore}{\texttt{He core}\xspace}
\newcommand{\COcore}{\texttt{C/O core}\xspace}

\usepackage{amsmath}
\usepackage{multirow}

\shorttitle{Double Neutron Stars at Solar Metallicity}
\shortauthors{Chattaraj et al. 2025}

\begin{document}

\title{Forming Double Neutron Stars using Detailed Binary Evolution Models with POSYDON: Comparison to the Galactic Systems}

\author[0000-0002-6064-388X]{Abhishek Chattaraj}\email{a.chattaraj@ufl.edu}
\affiliation{Department of Physics, University of Florida, 2001 Museum Rd, Gainesville, FL 32611, USA}

\author[0000-0001-5261-3923]{Jeff J. Andrews}
\affiliation{Department of Physics, University of Florida, 2001 Museum Rd, Gainesville, FL 32611, USA}
\affiliation{Institute for Fundamental Theory, 2001 Museum Rd, Gainesville, FL 32611, USA}

\author[0000-0002-3439-0321]{Simone\,S.\,Bavera}
\affiliation{Département d’Astronomie, Université de Genève, Chemin Pegasi 51, CH-1290 Versoix, Switzerland}
\affiliation{Gravitational Wave Science Center (GWSC), Université de Genève, CH1211 Geneva, Switzerland}

\author[0000-0002-6842-3021]{Max\,Briel}
\affiliation{Département d’Astronomie, Université de Genève, Chemin Pegasi 51, CH-1290 Versoix, Switzerland}
\affiliation{Gravitational Wave Science Center (GWSC), Université de Genève, CH1211 Geneva, Switzerland}

\author[0000-0001-5867-5033]{Debatri Chattopadhyay}
\affiliation{Center for Interdisciplinary Exploration and Research in Astrophysics (CIERA), Northwestern University, 1800 Sherman Ave, Evanston, IL 60201, USA}
\affiliation{NSF-Simons AI Institute for the Sky (SkAI),172 E. Chestnut St., Chicago, IL 60611, USA}

\author[0000-0003-1474-1523]{Tassos\,Fragos}
\affiliation{Département d’Astronomie, Université de Genève, Chemin Pegasi 51, CH-1290 Versoix, Switzerland}
\affiliation{Gravitational Wave Science Center (GWSC), Université de Genève, CH1211 Geneva, Switzerland}

\author[0000-0001-6692-6410]{Seth\,Gossage}
\affiliation{Center for Interdisciplinary Exploration and Research in Astrophysics (CIERA), Northwestern University, 1800 Sherman Ave, Evanston, IL 60201, USA}
\affiliation{NSF-Simons AI Institute for the Sky (SkAI),172 E. Chestnut St., Chicago, IL 60611, USA}

\author[0000-0001-9236-5469]{Vicky\,Kalogera}
\affiliation{Center for Interdisciplinary Exploration and Research in Astrophysics (CIERA), Northwestern University, 1800 Sherman Ave, Evanston, IL 60201, USA}
\affiliation{NSF-Simons AI Institute for the Sky (SkAI),172 E. Chestnut St., Chicago, IL 60611, USA}
\affiliation{Department of Physics and Astronomy, Northwestern University, 2145 Sheridan Road, Evanston, IL 60208, USA}

\author[0000-0003-3684-964X]{Konstantinos\,Kovlakas}
\affiliation{Institute of Space Sciences (ICE, CSIC), Campus UAB, Carrer de Magrans, 08193 Barcelona, Spain}
\affiliation{Institut d'Estudis Espacials de Catalunya (IEEC),  Edifici RDIT, Campus UPC, 08860 Castelldefels (Barcelona), Spain}

\author[0000-0001-9331-0400]{Matthias\,U.\,Kruckow}
\affiliation{Département d’Astronomie, Université de Genève, Chemin Pegasi 51, CH-1290 Versoix, Switzerland}
\affiliation{Gravitational Wave Science Center (GWSC), Université de Genève, CH1211 Geneva, Switzerland}

\author[0000-0002-8883-3351]{Camille\,Liotine}
\affiliation{Center for Interdisciplinary Exploration and Research in Astrophysics (CIERA), Northwestern University, 1800 Sherman Ave, Evanston, IL 60201, USA}
\affiliation{Department of Physics and Astronomy, Northwestern University, 2145 Sheridan Road, Evanston, IL 60208, USA}

\author[0000-0003-4474-6528]{Kyle\,A.\,Rocha}
\affiliation{Center for Interdisciplinary Exploration and Research in Astrophysics (CIERA), Northwestern University, 1800 Sherman Ave, Evanston, IL 60201, USA}
\affiliation{NSF-Simons AI Institute for the Sky (SkAI),172 E. Chestnut St., Chicago, IL 60611, USA}
\affiliation{Department of Physics and Astronomy, Northwestern University, 2145 Sheridan Road, Evanston, IL 60208, USA}

\author[0000-0003-1749-6295]{Philipp\,M.\,Srivastava}
\affiliation{Center for Interdisciplinary Exploration and Research in Astrophysics (CIERA), Northwestern University, 1800 Sherman Ave, Evanston, IL 60201, USA}
\affiliation{NSF-Simons AI Institute for the Sky (SkAI),172 E. Chestnut St., Chicago, IL 60611, USA}
\affiliation{Electrical and Computer Engineering, Northwestern University, 2145 Sheridan Road, Evanston, IL 60208, USA}

\author[0000-0001-9037-6180]{Meng\,Sun}
\affiliation{Center for Interdisciplinary Exploration and Research in Astrophysics (CIERA), Northwestern University, 1800 Sherman Ave, Evanston, IL 60201, USA}

\author[0000-0003-0420-2067]{Elizabeth\,Teng}
\affiliation{Center for Interdisciplinary Exploration and Research in Astrophysics (CIERA), Northwestern University, 1800 Sherman Ave, Evanston, IL 60201, USA}
\affiliation{NSF-Simons AI Institute for the Sky (SkAI),172 E. Chestnut St., Chicago, IL 60611, USA}
\affiliation{Department of Physics and Astronomy, Northwestern University, 2145 Sheridan Road, Evanston, IL 60208, USA}

\author[0000-0002-0031-3029]{Zepei\,Xing}
\affiliation{Département d’Astronomie, Université de Genève, Chemin Pegasi 51, CH-1290 Versoix, Switzerland}
\affiliation{Gravitational Wave Science Center (GWSC), Université de Genève, CH1211 Geneva, Switzerland}

\author[0000-0002-7464-498X]{Emmanouil\,Zapartas}
\affiliation{Département d’Astronomie, Université de Genève, Chemin Pegasi 51, CH-1290 Versoix, Switzerland}
\affiliation{Institute of Astrophysics, Foundation for Research and Technology-Hellas, GR-71110 Heraklion, Greece}

\begin{abstract}
With over two dozen detections in the Milky Way, double neutron stars (DNSs) provide a unique window into massive binary evolution. We use the \posydon binary population synthesis code to model DNS populations and compare them to the observed Galactic sample. By tracing their origins to underlying single and binary star physics, we place constraints on the detailed evolutionary stages leading to DNS formation. Our study reveals a bifurcation within the well-known common envelope (CE) formation channel for DNSs, which naturally explains an observed split in the orbital periods of the Galactic systems. The two subchannels are defined by whether the donor star has a helium core (Case B mass transfer) or a carbon-oxygen core (Case C) at the onset of the CE, with only the helium core systems eventually merging due to gravitational wave-modulated orbital decay. 
We find that across different treatments of the CE phase, the formation of DNSs through both subchannels requires either a generous core definition of $\simeq$ 30\% H-fraction or a high CE ejection efficiency of $\alpha_{\rm CE}\gtrsim1.2$. By testing different supernova kick velocity models, we find that galactic DNSs are best reproduced using a prescription that favors low velocity kicks ($\lesssim 50 \, \rm km/s$), in agreement with previous studies. Furthermore, our models indicate that merging DNSs are born from a stripped progenitor with a median pre-supernova envelope mass $\sim$ 0.2$\Msun$. 
Our results highlight the value of detailed evolutionary models for improving our understanding of exotic binary star formation.
\end{abstract}


\section{Introduction}\label{sec:intro}

Double neutron stars (DNSs) offer unique environments to explore the limits of fundamental physics, providing valuable insights into stellar evolution, supernova mechanisms, and the nature of matter under extreme conditions. 
NSs in binaries can be observed across a range of electromagnetic signatures---as radio pulsars, X-ray binaries, or through gravitational waves (GWs) emitted during mergers, which may be accompanied by kilonovae or short gamma-ray bursts, bridging two worlds of observational astrophysics. The preeminent example of this was the first detected DNS merger event, GW170817 \citep{LIGOGW170817+2017}, which was followed by a short gamma-ray burst \citep{LIGOGW170817+sGRB} and a kilonova \citep{LIGOkilonova2017}, giving rise to a new era of multimessenger astronomy. 
During a later observing run, a second DNS merger event was detected, GW190425 \citep{LIGOGW190425+2020}, although without an electromagnetic counterpart.

Among LIGO sources, the DNS population maintains a unique advantage in that there exists a meaningful sample of known systems within the Milky Way (MW). Although other formation mechanisms exist \citep[e.g., formation in triple systems or dynamically within globular clusters;][]{2019ApJ...883...23H, 2020ApJ...888L..10Y}, it is widely believed that most DNSs are the final products of the complex evolution of isolated massive binaries over millions of years \citep{Bhattacharya&vandenHeuvel1991}. Starting with the discovery of the Hulse-Taylor binary \citep{Hulse&Taylor1975}, we now know over two dozen DNSs in the Galaxy, roughly half of which will merge within a Hubble time due to gravitational wave radiation \citep{Tauris+2017}.

The isolated evolution of massive binaries leading to the formation of DNS systems has been extensively discussed in prior studies \citep[][]{Zwart&Yungelson1998, Belczynski+2002, Oslowski+2011, Andrews+2015, Tauris+2017, Alejandro+2018, Kruckow+2018, Chattopadhyay+2020,Broekgaarden:2021efa,Chu+2022, Cecilia+2023, Deng+2024}. Here, we provide a broad overview of the typical evolutionary scenario. The primary formation channel involves a common envelope (CE) phase \citep{Ostriker1973, Paczynski+1976} of the first-born neutron star (NS) with its H-rich companion, which is crucial for hardening the binary. This is followed by an additional round of mass transfer (MT) with the helium giant \citep{Ivanova+2003, Dewi&Pols2003}, rejuvenating the first-born NS and creating a millisecond pulsar \citep[MSP;][]{Alpar+1982, Radhakrishnan&Srinivasan1982, Bhattacharya&vandenHeuvel1991}. Throughout its evolution, the binary must survive two supernova (SN) explosions to remain bound. These explosions can impart a combination of a Blauuw kick due to instantaneous mass loss during core collapse \citep{Blaauw1961} and a natal kick directly to the nascent NS due to asymmetries in the core collapse process \citep{Lyne&Lorimer1994}. Although multiple pieces of evidence have demonstrated that isolated radio pulsars receive natal kicks in excess of several hundred km/s \citep{Gunn&Ostriker1970, Lyne&Lorimer1994, Hobbs+2005, Chatterjee+2005, Verbunt+2017, Disberg&Mandel2025}, it is not clear that the same process applies to neutron stars in binaries \citep{Willcox+2021}. While some DNS systems, such as B1534$+$12 and B1913$+$16, have been shown to form with large kick velocities \citep{Fryer&Kalogera1997, Willems+2004, Thorsett+2005, Wong+2010}, it has been recognized that at least some systems are born with significantly lower kicks \citep{Podsiadlowski+2004, Heuvel2004, Willems&Kalogera2004, Willems+2004, Stairs+2006, Beniamini&Piran2016, O'Doherty+2023}. If the binary remains bound after both core collapse events, which is significantly more likely for weaker kicks but also depends on the kick direction, DNSs in sufficiently tight orbits will eventually merge in a Hubble time, emitting gravitational radiation \citep{LIGOGW170817+2017, LIGOGW190425+2020, LIGONSBH+2021}, potentially triggering a short gamma-ray burst \citep[GRB;][]{Eichler+1989, Berger2014, Goldstein+2017, Hotokezaka+2018, Murase+2018} or even a long GRB \citep{Rastinejad+2022, Levan+2023, Gottlieb+2023}, producing a kilonova \citep{Li&Paczynski1998, Kulkarni2005, Metzger+2010, Tanvir+2013, Tanaka&Hotokezaka2013, LIGOkilonova2017, Troja+2017, Kasen+2017, Nicholl+2017, Soares-Santos+2017, Cowperthwaite+2017, Smartt+2017, Metzger2019}, and leading to the $r$-process enrichment of galaxies \citep{Lattimer&Schramm1974, Eichler+1989, Mennekens&Vanbeveren2014, Cote+2018, Safarzadeh+2019, Kobayashi+2023, Holmbeck&Andrews2023, Chen+2024b}.

\begin{figure}
\includegraphics[width=\linewidth]{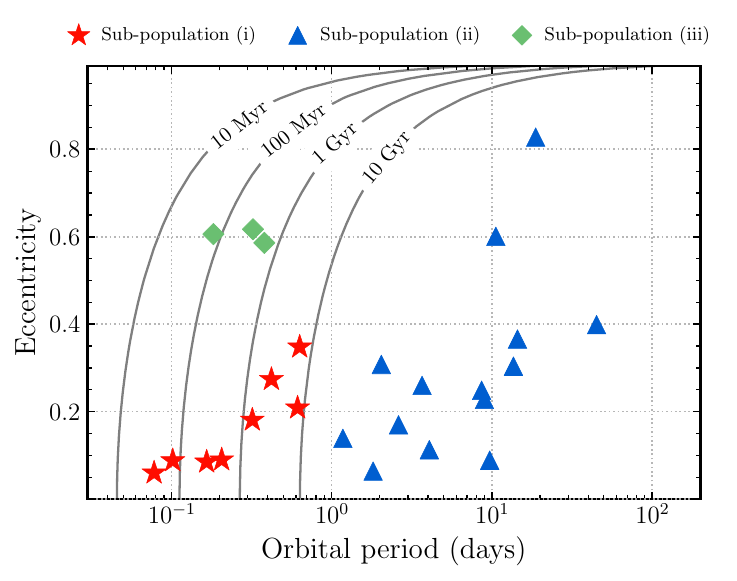}
\caption{Distribution of the 25 currently known or suspected DNS binaries in the MW field (Table~\ref{tab:DNS_observations}), categorized into three sub-populations following \citet{Andrews&Mandel2019}. The sub-populations, which likewise cluster in $P-\dot{P}$ space (see Figure \ref{fig:Obs_P-Pdot}), has remained broadly consistent with the known population today even after including six additional systems since the claim by \citet{Andrews&Mandel2019}. For reference, lines of constant gravitational wave merger times have been plotted for $1.4\,\Msun$ NS components.}
\vspace{1em}
\label{fig:Obs_Porb-e}
\centering
\includegraphics[width=\linewidth]{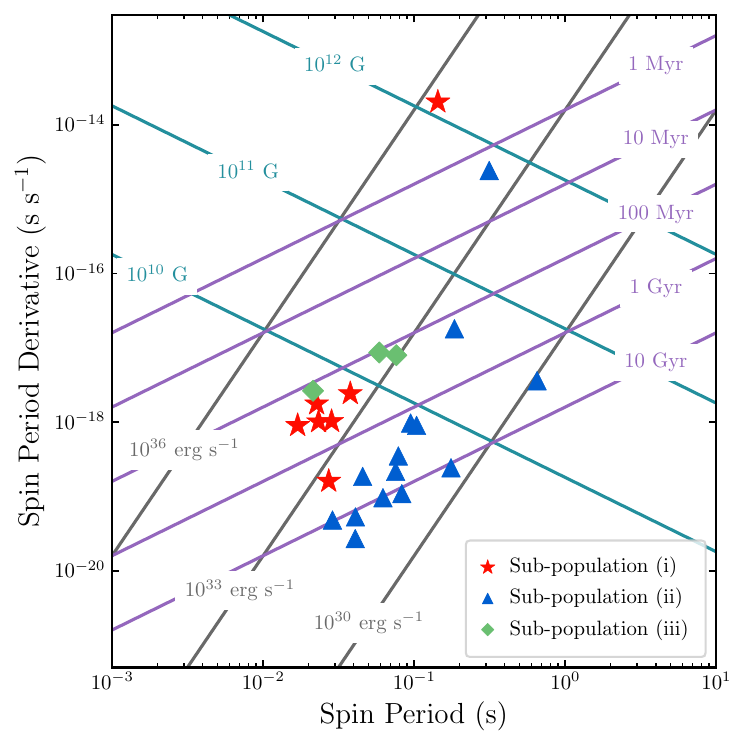}
\caption{The $P-\dot{P}$ distribution of the observed pulsars show a clustering consistent with the three DNS sub-populations in Figure \ref{fig:Obs_Porb-e}. For reference, we plot lines of constant magnetic field (teal), spin-down luminosity (gray), and characteristic age (purple). These lines are calculated assuming dipole radiation from a  $1.4\,\Msun$ NS with a radius of 12 km, a moment of inertia of $(2/5) M R^2$, and an inclination angle of $\pi/2$ between the magnetic field and spin axes.}
\label{fig:Obs_P-Pdot}
\end{figure}

\setlength{\tabcolsep}{10pt}
\begin{table*}[]
    \centering
      \begin{tabular}{lllllllll}
  \hline
  \hline
  \quad System & $P_{\rm orb}$ & $e$ & $P_{\rm spin}$ & $P_{\rm dot}$ & $\tau_{\rm c}$ & $M_{\rm p}$ & $M_{\rm c}$ & $M_{\rm sys}$ \\
   & (days) & & (ms) & (10$^{-18}$) & (Myr) & $(\mathrm{\Msun})$ & $(\mathrm{\Msun})$ & $(\mathrm{\Msun})$ \\
  \hline
  \multicolumn{6}{c}{\hspace{6cm} Sub-population (i)} \\
  \hline
  J0737$-$3039$^a$ & 0.102 & 0.088 & 22.7 & 1.76 & 204 & 1.338 & 1.249 & 2.587 \\
  B1534$+$12$^b$ & 0.421 & 0.274 & 37.9 & 2.42 & 248 & 1.3330 & 1.345 & 2.678 \\
  J1756$-$2251$^c$ & 0.320 & 0.181 & 28.5 & 1.02 & 443 & 1.341 & 1.230 & 2.570 \\
  J1906$+$0746$^{d *}$ & 0.166 & 0.085 & 144.1 & 20300 & 0.1 & 1.291 & 1.322 & 2.613 \\
  J1913$+$1102$^e$ & 0.206 & 0.090 & 27.3 & 0.16 & 2687 & 1.62 & 1.27 & 2.889 \\
  J1946$+$2052$^f$ & 0.078 & 0.06 & 17.0 & 0.9 & 299 & $<1.31$ & $>1.18$ & 2.50 \\
  J1208$-$5936$^g$ & 0.632 & 0.348 & 28.7 & $<0.04$ & $>10000$ & 1.26 & 1.32 & 2.586 \\
  J1846$-$0513$^h$ & 0.613 & 0.209 & 23.4 & 1.01 & 368 & $\leq$ 1.3455 & $\geq$ 1.2845 & 2.629 \\
  \hline
  \multicolumn{6}{c}{\hspace{6cm} Sub-population (ii)} \\
  \hline
 J0453$+$1559$^i$ & 4.072 & 0.113 & 45.8 & 0.186 & 3901 & 1.559 & 1.174 & 2.734 \\
  J1411$+$2551$^j$ & 2.616 & 0.1699 & 62.4 & 0.096 & 10342 & $<1.62$ & $>0.92$ & 2.538 \\
  J1518$+$4904$^k$ & 8.634 & 0.249 & 40.9 & 0.027 & 23824 & 1.41 & 1.31 & 2.718 \\
  J1753$-$2240$^l$ & 13.638 & 0.304 & 95.1 & 0.970 & 1553 & -- & $>0.4875$ & -- \\
  J1755$-$2550$^{m *}$ & 9.696 & 0.089 & 315.2 & 2430 & 2 & -- & $>0.39$ & -- \\
  J1811$-$1736$^n$ & 18.779 & 0.828 & 104.2 & 0.901 & 1832 & $<1.64$ & $>0.93$ & 2.57 \\
  J1829$+$2456$^o$ & 1.176 & 0.139 & 41.0 & 0.053 & 12373 & 1.306 & 1.299 & 2.605 \\
  J1930$-$1852$^p$ & 45.060 & 0.399 & 185.5 & 18.0 & 163 & $<1.32$ & $>1.30$ & 2.59 \\
  J1325$-$6253$^q$ & 1.816 & 0.064 & 28.9 & 0.048 & 9259 & $<1.59$ & $>0.98$ & 2.57 \\
  J1759$+$5036$^r$ & 2.043 & 0.308 & 176.0 & 0.243 & 11500 & $<1.9194$ & $>0.7006$ & 2.62 \\
  J1018$-$1523$^s$ & 8.984 & 0.228 & 83.2 & 0.109 & 12115 & $<1.1$ & $>1.16$ & 2.3 \\
  J2150$+$3427$^{t *}$ & 10.592 & 0.601 & 654 & 3.60 & 2883 & $<1.67$ & $>0.98$ & 2.59 \\ 
  J1901$+$0658$^u$ & 14.455 & 0.366 & 75.7 & 0.217 & 5500 & $<1.68$ & $>1.11$ & 2.79 \\
  J1155$-$6529$^v$ & 3.668 & 0.260 & 78.9 & 0.35 & 3574 & -- & -- & -- \\
  \hline
  \multicolumn{6}{c}{\hspace{6cm} Sub-population (iii)} \\
  \hline
  J0509$+$3801$^w$ & 0.380 & 0.586 & 76.5 & 7.93 & 153 & 1.46 & 1.34 & 2.805 \\
  J1757$-$1854$^x$ & 0.183 & 0.606 & 21.5 & 2.63 & 130 & 1.3412 & 1.3917 & 2.733 \\
  B1913$+$16$^y$ & 0.323 & 0.617 & 59.0 & 8.63 & 108 & 1.4398 & 1.3886 & 2.828 \\
  \hline
  \hline
  \end{tabular}
     \caption{
    Properties of the currently known sample of DNSs in the Milky Way field.
    $^*$Unconfirmed DNS; a NS--WD solution could be possible. Characteristic ages are calculated assuming an angle of $\pi/2$ between the magnetic field and spin axes.}
    \begin{flushleft}
    $^a$ \citet{Burgay+03}, \citet{Kramer+06}, \citet{Ferdman+2013}, \citet{Kramer+2021}; $^b$ \citet{Stairs+2002};  $^c$ \citet{Faulkner+2005}, \citet{Fonseca+2014}, \citet{Ferdman+2014}; $^d$ \citet{Lorimer+2006}, \citet{Leeuwen2015}; $^e$ \citet{Lazarus2016}, \citet{Ferdman+2020}; $^f$ \citet{Stovall+2018}; $^g$ \citet{Bernadich+2023}; $^h$ \citet{Zhao+2024}; $^i$ \citet{Martinez+2015}; $^j$ \citet{Martinez+2017}; $^k$ \citet{Janssen+2008}; $^l$ \citet{Keith+2009}; $^m$ \citet{Ng+2015}, \citet{Ng+2018}; $^n$ \citet{Lyne+2000}, \citet{Corongiu+2007};
    $^o$ \citet{Champion+2004}, \citet{Haniewicz+2021}; $^p$ \citet{Swiggum+2015}; $^q$ \citet{Sengar+2022}; $^r$ \citet{Agazie+2021}; $^s$ \citet{Swiggum+2023}; $^t$ \citet{Wu+2023}; $^u$ \citet{Han+2021}, \citet{Su+2024}; $^v$ \citet{Padmanabh+2023};
    $^w$ \citet{Lynch+2018}; $^x$ \citet{Cameron+2018}, \citet{Cameron+2023}; $^y$ \citet{Hulse&Taylor1975}, \citet{Kramer1998}, \citet{Weisberg+2010}, \citet{Weisberg&Huang2016}.
    \end{flushleft}
    \label{tab:DNS_observations}
\end{table*}

The best observational constraints on DNS formation come from the 25 DNS binaries identified in the Galactic Field (although it is possible that a few of these may be NS--WD binaries rather than DNSs; see Table \ref{tab:DNS_observations}). 
We exclude from this list the additional 4 candidate DNS binaries in globular clusters presumably formed through dynamical interactions: B2127$+$11C in M15 \citep{Anderson+1990}, J1807$-$2500 in NGC 6544 \citep{Lynch+2012}, J0514$-$4002 in NGC 1851 \citep{Ridolfi+2019}, and J1748$-$2446ao in Terzan 5 \citep{2024A&A...686A.166P}. Our focus in this paper will be the entries presented in Table \ref{tab:DNS_observations}, all of which reside in the MW disk and are therefore expected to have originated from isolated binary evolution. 

Using a smaller sample of 17 DNSs, \cite{Andrews&Mandel2019} categorized these systems into three sub-populations based on their orbital periods and eccentricities. Sub-population (i) comprises of short period systems ($P_{\rm orb} \sim 0.08 - 0.6$ days) with low eccentricities ($e < 0.35$) which will merge within a Hubble time. Sub-population (ii) includes wide binaries ($P_{\rm orb} \sim 1 - 19$ days) that will not merge in a Hubble time. Sub-population (iii) consists of DNSs with short periods ($P_{\rm orb} < 0.4$ days) and eccentricities tightly clustered around $e \simeq 0.6$. \cite{Andrews&Mandel2019} proposed that this last sub-population is consistent with dynamical assembly in globular clusters followed by ejection into the field, although isolated evolution could be responsible with fine tuning. These sub-populations likewise cluster in the diagram of spin-period and its derivative, with sub-population (i) typically having shorter spin periods compared to sub-population (ii).
With six additional detections in the MW field in the intervening years since \citet{Andrews&Mandel2019}, the current sample shown in Figures \ref{fig:Obs_Porb-e} and \ref{fig:Obs_P-Pdot} continue to exhibit three clusters in the $P_{\rm orb} - e$ and $P-\dot{P}$ plane.

While DNSs have been the focus of multiple studies in the past, two developments justify the current work: First, the observational sample has expanded greatly in recent years, averaging one new detection per year for the past decade. Second, while previous population synthesis studies of DNSs have relied on fitting formula or libraries of single star evolution tracks, the recent release of \posydon \citep[{\tt PO}pulation {\tt SY}nthesis with {\tt D}etailed binary-evolution simulati{\tt ON}s, see][]{POSYDON+2023, Andrews+2025} offers a fresh approach on modeling binary populations. \posydon  employs fully self-consistent and detailed binary and single star evolution, avoiding many of the limitations of rapid binary population synthesis (BPS) codes.  
We provide an overview of \posydon and elaborate on our methodology for generating massive binary populations in Section \ref{sec:method}. In Section \ref{sec:grids}, we delve into the detailed evolutionary channels that lead to DNS formation as indicated by \posydon, while Section \ref{sec:ce} presents constraints on our understanding of the CE phase. In Section \ref{sec:casebb}, we quantify the mass accretion process and its impact on pulsar recycling, followed by Section \ref{sec:supernova} where we present results from supernova modeling including natal kicks and DNS masses. Section \ref{sec:merger_rates} explores the merger rates derived from our population models. We then compare our findings to previous studies in Section \ref{sec:comparison}. Finally, we summarize our results and discuss future prospects in Section \ref{sec:conclusions}.

\section{Method} \label{sec:method}

\setlength{\tabcolsep}{9.5pt}
\begin{table*}[]
\centering
\caption{Summary of the model populations and their key properties used in this study. The models are grouped by the three common envelope treatments (see Section \ref{subsec:model_variations}). The columns list the model name, the energy conversion efficiency for common envelope ejection $\alpha_{\rm CE}$, the adopted hydrogen fraction defining the core-envelope boundary, the supernova model, and the natal kick prescription. For each population, we report the fraction of Fe core-collapse supernovae (while the rest undergo electron-capture supernovae) between the two supernova events, and the local ($z=0$) merger rate density $\mathcal{R}$ (in $\mathrm{Gpc^{-3} \ yr^{-1}}$).
}

\begin{tabular}{cccccccc}
\hline
\hline
\multirow{2}{*}{Model Name} & \multicolumn{2}{c}{$\textrm{CE parameters}$} & \multirow{2}{*}{SN model$^a$} & \multirow{2}{*}{Kick model$^b$}
& \multicolumn{2}{c}{\textrm{\% CCSN}} & \multicolumn{1}{c}{$\mathcal{R}$ ($z$=0)} \\
\cline{2-3}
\cline{6-7}
& $\mathrm{\alpha_{CE}}$ & H-fraction & & & SN1 & SN2 & [\textrm{$\rm Gpc^{-3} ~ yr^{-1}$}] \\
\noalign{\smallskip}
\hline
  \multicolumn{6}{c}{\hspace{3.5cm} \texttt{Two\_Phases\_StableMT}} \\
  \hline
\texttt{MODEL01} & 1 & 30\% & S16 & Hobbs-Maxwellian & 70\% & 53\% & 1.19 \\
\texttt{MODEL02} & 1 & 30\% & S16 & AsymEj & 99\% & 81\% & 54.94 \\
\texttt{MODEL03} & 1 & 30\% & PS20 & Hobbs-Maxwellian & 76\% & 65\% & 1.75 \\
\texttt{MODEL04} & 1 & 30\% & PS20 & AsymEj & 99\% & 81\% & 36.16 \\
\texttt{MODEL05} & 1 & 30\% & F12-delayed & Hobbs-Maxwellian & 89\% & 78\% & 3.43 \\
\texttt{MODEL06} & 1 & 30\% & F12-delayed & AsymEj & 99\% & 86\% & 153.67 \\
\texttt{MODEL07} & 1 & 30\% & F12-rapid & Hobbs-Maxwellian & 76\% & 68\% & 1.35 \\
\texttt{MODEL08} & 1 & 30\% & F12-rapid & AsymEj & 99\% & 83\% & 46.16 \\
\texttt{MODEL09} & 1 & 30\% & S16 & Low Maxwellian$^{*}$ & 99\% & 87\% & 359.43 \\
\texttt{MODEL10} & 1 & 1\% & S16 & AsymEj & 99\% & 82\% & 12.33 \\
\texttt{MODEL11} & 3 & 30\% & S16 & Hobbs-Maxwellian & 89\% & 55\% & 5.22 \\
\texttt{MODEL12} & 3 & 10\% & S16 & AsymEj & 99\% & 81\% & 155.85 \\
\texttt{MODEL13} & 5 & 1\% & S16 & AsymEj & 99\% & 80\% & 700.54 \\
\texttt{MODEL14} & 5 & 1\% & S16 & Linear & 99\% & 78\% & 117.69 \\
\hline
  \multicolumn{6}{c}{\hspace{3.5cm} \texttt{Two\_Phases\_Windloss}} \\
  \hline
\texttt{MODEL15} & 1 & 30\% & S16 & Hobbs-Maxwellian & 83\% & 68\% & 3.16 \\
\texttt{MODEL16} & 1 & 30\% & S16 & AsymEj & 99\% & 81\% & 135.82 \\
\texttt{MODEL17} & 3 & 10\% & S16 & AsymEj & 99\% & 82\% & 411.65 \\
\texttt{MODEL18} & 3 & 10\% & F12-delayed & Linear & 99\% & 88\% & 335.68 \\
\hline
  \multicolumn{6}{c}{\hspace{3.5cm} \texttt{One\_Phase}} \\
  \hline
\texttt{MODEL19} & 1 & 30\% & F12-delayed & Hobbs-Maxwellian & 75\% & 86\% & 7.52 \\
\texttt{MODEL20} & 1 & 30\% & PS20 & Hobbs-Maxwellian & 64\% & 82\% & 2.83 \\
\texttt{MODEL21} & 1 & 30\% & S16 & Hobbs-Maxwellian & 59\% & 83\% & 4.29 \\
\texttt{MODEL22} & 1 & 30\% & S16 & AsymEj & 98\% & 88\% & 276.61 \\
\texttt{MODEL23} & 3 & 10\% & F12-delayed & Linear & 97\% & 92\% & 275.64 \\
\texttt{MODEL24} & 3 & 10\% & S16 & AsymEj & 98\% & 87\% & 451.65 \\
\texttt{MODEL25} & 5 & 1\% & S16 & Linear & 99\% & 77\% & 117.96 \\
\noalign{\smallskip}
\hline 
\hline
\end{tabular}
\begin{flushleft}
$^a$S16: \citet{Sukhbold+2016}; PS20: \citet{Patton&Sukhbold2020}; F12: \cite{Fryer+2012}.\\
$^b$Hobbs-Maxwellian: \citet{Hobbs+2005}, $^{*}\sigma_{\rm CCSN} = 20\,{\rm km\,s^{-1}}$;
AsymEj: \citet{Janka2017};
Linear: \citet{Richards+2023}. 
\end{flushleft}
\label{tab:rates}
\end{table*}

\subsection{Overview of the binary population synthesis code \rm \posydon}
\label{subsec:POSYDON}

To investigate the formation and detailed evolution of DNSs in the Galactic Field, we utilize the state-of-the-art binary population synthesis suite \posydon \footnote{The version of the code is identified by the commit hash f4d68ed available at \href{https://github.com/POSYDON-code/POSYDON/}{https://github.com/POSYDON-code/POSYDON/}. The version of the grids used in this study can be found here: \citet{https://doi.org/10.5281/zenodo.14205146}.} \citep{POSYDON+2023, Andrews+2025}, which incorporates detailed single and binary stellar evolution models based on the 1D stellar evolution code \mesa \citep{Paxton+2011, Paxton+2013, Paxton+2015, Paxton+2018, Paxton+2019, Jermyn+2023}. \posydon consists of five pre-computed stellar evolution grids, which includes two grids of single H-rich and He-rich stars (used to model non-interacting, detached binaries) and three binary star grids of: (i) two H-rich stars at zero-age main sequence (ZAMS), the \texttt{HMS-HMS} grid; (ii) a compact object (CO) with a H-rich companion at the onset of Roche Lobe Overflow (RLO), the \texttt{CO-HMS\_RLO} grid; and (iii) a CO with a He-rich companion at zero-age helium main sequence (ZAHeMS), the \texttt{CO-HeMS} grid. By applying classification and interpolation algorithms to the grids, \posydon calculates the evolution of a binary system with arbitrary initial conditions. On top of this infrastructure, we implement extensive on-the-fly calculations that incorporate essential physics such as magnetic braking, GW radiation, CE evolution, CO formation, SN kicks, etc. This framework allows for the simultaneous study of self-consistent evolution of the binary system from ZAMS to the merger of a double compact object (DCO), along with the internal structure of the individual stars. 
Below, we summarize the key physics assumptions we adopt in simulating our model populations.

\textit{Initializing our binary populations}: We draw masses of the primary\footnote{Throughout this work, primary star refers to the initially more massive star, and secondary refers to the initially less massive star regardless of the binary's evolutionary state.} star from the initial mass function (IMF) outlined in \cite{2001MNRAS.322..231K} within the range $\rm{M_1} \in [6.5, 150]$$\,\Msun$. The masses of the secondary component follow a flat distribution between [0.05, 1] in mass ratio $q$. While evidence suggests the mass ratio distribution of short-period, high-mass binaries is similar to uniform \citep{2017ApJS..230...15M}, as we show in Appendix \ref{sec:HMS_HMS}, most DNSs originate from binaries with ZAMS mass ratios close to unity. Therefore we expect that even modest variations from this distribution will have little impact on our results. We assume a binary fraction of unity for all our models. The initial orbit is circular with the orbital period for systems with $\rm{M_1}\leq15$$\Msun$ drawn from a flat distribution in $\log \rm{P}$ over the range $[10^{0.35},10^{3.5}]$ days. For $\rm{M_1}>15$$\Msun$, we adopt an extension to the \citet{Sana+2012} power-law distribution, as described in Appendix A of \citet{Bavera+2021}. We have tested alternative initial period distributions and found our results are robust to reasonable variations. Finally, we simulate a constant star formation history (SFH) by assigning random birth ages from a uniform distribution between 0 and 13.8 Gyr (a realistic SFH is used when calculating merger rates). While the Galactic SFH has evolved over cosmic time, most observed DNSs were formed within the past $\sim$1 Gyr during which the SFH has remained relatively uniform \citep[e.g.,][]{2015A&A...578A..87S}.

\textit{Mass transfer}: As the \posydon grids used in this study consist of detailed \mesa binary simulations, the rates of mass transfer due to RLO, as well as its impact on both the donor's and accretor's structure are determined in a self-consistent manner. Upon RLO, \posydon instantaneously circularizes the orbit at pericenter. 
While accretion onto a non-degenerate star is assumed to be conservative until the accretor is spun up to near break-up velocities, accretion onto a degenerate star is Eddington-limited. A more detailed description of mass transfer in \posydon can be found in Section 4.2 of \citet{POSYDON+2023}. 

\textit{Common Envelope}: A CE is triggered in \posydon if one of the following three main conditions are met: (i) the mass loss rate of the donor exceeds 0.1$\,\Msun\,\rm yr^{-1}$, (ii) RLO occurs via the outer L2 Lagrangian point, or (iii) the photon trapping radius of the CO accretor becomes comparable to its Roche lobe radius \citep[for additional details, see Section 4.2.4 in][]{POSYDON+2023}. Our default model in \posydon uses a two-step process to calculate the orbital evolution of a binary through a CE. The first step is an adaptation of the $\alpha_{\rm CE}\,$--$\,\lambda_{\rm CE}$ treatment \citep{Webbink1984, Livio&Soker1988}, where the parameter $\alpha_{\rm CE}$ represents the efficiency of utilizing orbital energy to unbind the envelope ($\alpha_{\rm CE} \, \Delta E_{\rm orb} = \Delta E_{\rm bind}$). Motivated by arguments from \citet{Ivanova2011} and simulations from \citet{Fragos+2019} and \citet{Marchant+2021}, we assume the system detaches (provided there is enough orbital energy available) with a thin layer of the envelope still surrounding the core. In our default model, we adopt a stopping point for this first phase defined by a $30\%$ H-mass fraction, although we explore two alternative definitions, at $1\%$ and $10\%$ H-mass fraction. The parameter $\lambda_{\rm CE}$, which sets the envelope binding energy and therefore determines which systems merge within a CE, is calculated from the star's structure adopting one of these boundaries. We assume a second phase of stable mass transfer ensues in which the mass of the companion is reduced down to its 1\% H-mass fraction boundary, assuming that the remaining thin envelope is lost from the system with the specific angular momentum of the accretor, with the orbit adjusted accordingly. This mass transfer phase is assumed to be stable, based off simulation results from \citet{Fragos+2019}.

We note that we calculate the CE
including the gravitational binding energy and internal energy of the envelope, and subtract the recombination energy contribution to the internal energy.
While previous studies using hydrodynamic simulations \citep{Ricker&Taam2008, Chamandy+2018} or ``wind tunnel" models \citep[e.g.,][]{Macleod&Ramirez-Ruiz2015, Macleod+2017} show that accretion during CE is possible, potentially at super-Eddington rates; in this work, for simplicity we have assumed no mass accretion onto the NS during the CE phase. Assuming neither star overfills its Roche lobe during either of these phases, the system survives the common envelope, albeit with a greatly reduced orbital separation. In subsequent evolution we treat the donor's remnant core as a He-star. 

\textit{Supernova and neutron star mass}: We follow the evolution of each star until core carbon exhaustion. Thereupon, we determine the SN outcome using the core-collapse prescriptions from \cite{Patton&Sukhbold2020}, \cite{Sukhbold+2016}, or the rapid and delayed models from \cite{Fryer+2012}. For a narrow range of core masses, a SN will occur due to electron captures within a degenerate oxygen-neon-magnesium (ONeMg) core (ECSN) rather than through typical Fe-core collapse \citep[CCSN;][]{Nomoto1984}. We determine whether a star undergoes ECSN based on its pre-SN carbon-oxygen (C/O) core mass, which must fall in the range $1.37\,\Msun \lesssim M_\mathrm{C/O\mbox{-}core} \lesssim 1.43\,\Msun$ \citep{Tauris+2015}.
The remnant baryonic mass $M_\mathrm{rembar}$, is then converted into the gravitational mass following \citet{Lattimer&Yahil}:
\begin{equation}
     M_\mathrm{grav} =
     \begin{cases}
     \dfrac{20}{3}\left(\sqrt{1 + 0.3 \dfrac{M_\mathrm{rembar}}{\Msun}} - 1\right){}\Msun,\, \\ \qquad \qquad \qquad
     M_\mathrm{rembar}-M_\mathrm{grav} < 0.5 \, \Msun \\
     M_\mathrm{rembar} - 0.5{}\Msun, 
     \qquad \qquad \text{otherwise}.
   \end{cases}
   \label{eq:rembar}
\end{equation}

Collapsing stars with $M_{\rm grav} < 2.5 \,\Msun$ are assigned to become NSs, while more massive stars become BHs, with our models setting the maximum mass for a NS to $M_{\rm NS}^{\rm max}=2.5\,\Msun$ \citep[see e.g.,][]{LIGOScientific:2020zkf}. Stars with core masses below the lower limits of ECSN evolve into WDs.
We assume no fallback onto the newborn NS except for
NSs born out of CCSN within either of the \cite{Fryer+2012} prescriptions. 

\textit{Natal kicks}: We draw the magnitude of the natal kick velocity $v_{\rm kick}$ randomly from a Maxwellian distribution with different dispersion values to distinguish between CCSN and ECSN scenarios: $\sigma_\mathrm{CCSN}=265\,\mathrm{km\,s^{-1}}$ \citep[based on single-star pulsar velocities from][]{Hobbs+2005} and $\sigma_\mathrm{ECSN}=20\,\mathrm{km\,s^{-1}}$ \citep[based on the argument that ECSN ought to be ``prompt'', leading to a much smaller kick velocity;][]{Podsiadlowski+2004}. Additionally, we test a model population with a low dispersion value of $\sigma_\mathrm{CCSN}=20\,\mathrm{km\,s^{-1}}$.

\textit{Merger rates}: To calculate DNS merger rates, we follow the procedure built into \posydon which we briefly describe here. We adopt a redshift and metallicity dependent star formation rate (SFR) based on the \texttt{IllustrisTNG} cosmological simulation suite \citep{Nelson+2019}. Focusing on star formation within the metallicity range \([0.5, 2]\,Z_\odot\), we convolve the resulting SFR density with our synthetic DNS population to report the intrinsic local (\(z = 0\)) merger rate density of DNSs formed at Solar metallicity. For a detailed calculation, see \citet{Bavera+2022} and \citet{Andrews+2025}.

\subsection{Model variations}
\label{subsec:model_variations}

We address the uncertainties in binary evolution by varying our prescriptions related to our common envelope treatment and our core-collapse treatment, including natal kicks, based on insights from prior studies highlighting these processes as particularly impactful \citep{Oslowski+2011, Andrews+2015, Kruckow+2018, Alejandro+2018, Cecilia+2023, Deng+2024}.

\textit{Variations to our Common Envelope Treatment}: For the first, $\alpha_{\rm CE}\,$--$\,\lambda_{\rm CE}$ phase of our CE prescription, we start with $\alpha_{\rm CE} = 1.0$, while exploring values below and above unity. We also experiment with varying definitions of the donor's core-envelope boundary, thereby altering the binding energy content of its envelope and influencing which systems survive a CE. In case of a successful ejection, we also vary our treatment of the second phase of our CE which removes the remaining thin H-envelope surrounding the donor's core. In addition to our default model which is meant to account for stable mass transfer of the remaining envelope onto the companion, we also include a model in which no second phase occurs, and whatever remaining H left over is treated as part of the He core, and a model in which the remaining thin envelope is emitted as a stellar wind, in which case the remaining thin envelope is lost with the specific angular momentum of the donor. Based on the above description, we list the following CE models used in this study:

(i) \texttt{One\_Phase}: No post-CE mass loss.

(ii) \texttt{Two\_Phases\_StableMT}: Post-CE mass loss with orbital shrinkage via stable, non-conservative MT.

(iii) \texttt{Two\_Phases\_Windloss}: Post-CE mass loss with orbital shrinkage via stellar winds.

\vspace{2pt}

\textit{Variations to our Core Collapse Treatment}: Stars that lose both their hydrogen and helium envelopes collapse in an ultra-stripped supernova (USSN) and are expected to have low ejecta masses \citep{Tauris+2013, Tauris+2015}, leading to considerably lower kick magnitudes \citep{Janka2012, Janka2017}. Moreover, at least some DNSs have been suggested to be formed with significantly lower kicks \citep{Podsiadlowski+2004, Heuvel2004, Willems+2004, Willems&Kalogera2004, Stairs+2006, Wong+2010, Beniamini&Piran2016, O'Doherty+2023}.
All of these factors suggest that the conventional Maxwellian kick distribution (with $\sigma_{\rm CCSN} = 265$ km s$^{-1}$) may not be accurate for modeling NS kicks in binary systems \citep{Bray&Eldridge2016}. Hence, in this work, we incorporate two additional models, which we term as ``physics-based kick prescriptions'' that apply natal kicks based on two different physics-informed approaches. We apply these prescriptions to stars collapsing in both a core-collapse SN and an ECSN.

\vspace{2pt}

\begin{figure*}[ht!]
\includegraphics[width=1.00\linewidth]{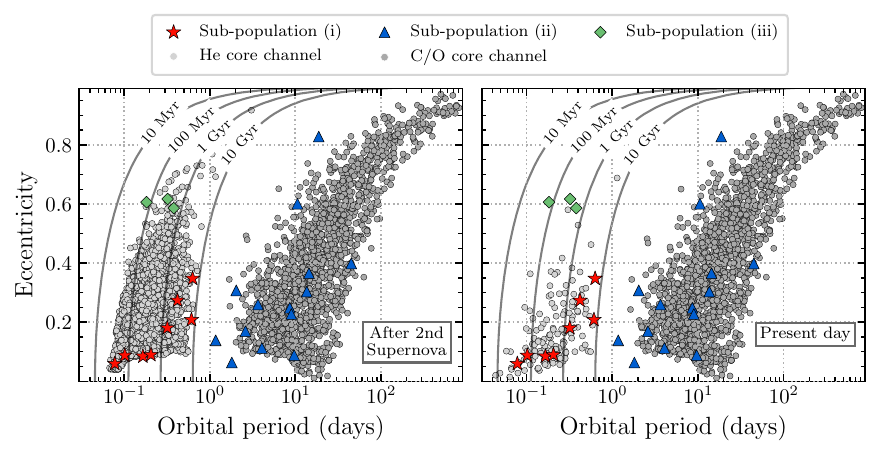}
\caption{Distribution of DNSs from \texttt{MODEL12} (see Table \ref{tab:rates}) on the $P_{\rm orb} - e$ plane. The left panel shows the distribution of simulated DNSs at birth, i.e. right after the second supernova, while the right panel depicts their distribution at the present day. The bifurcation in orbital periods represented by the two shades of gray markers indicate distinct sub-channels leading to DNS formation, one that is based on the evolutionary state of the H-rich donor at the onset of RLO with the first-born NS. While the light gray markers denote the binaries where mass transfer was initiated by a He-core donor (the \Hecore channel), the dark gray markers represent evolved donors with a carbon-oxygen core (\COcore channel). Additionally, we overplot the present day distribution of observed DNSs in the MW field (Table \ref{tab:DNS_observations}), and lines of constant merger times due to gravitational wave emission assuming a DNS pair with 1.4\,\Msun\ components. For a guide on comparing our simulated systems with the observations in the two panels, refer to text in Section \ref{subsec:comparison_to_observed_sample}.}
\label{fig:preferred_model}
\end{figure*}

(i) AsymEj (Asymmetric Ejecta) kicks: In this model, we determine the natal kick velocities imparted to the NS based on the gravitational tug-boat mechanism during asymmteric core-collapse \citep{Janka2017}. We apply an equivalent expression derived from Eq. 11 in \citet{Janka2017}, representing the SN explosion energy in terms of the ejected mass $M_\mathrm{ej}$, which is provided by \posydon for each SN event in a binary:
\begin{eqnarray}
v_\mathrm{kick} 
&=& 21\,\mathrm{km\,s}^{-1}
\left(\epsilon_5\,f_\mathrm{kin}\,\beta_\nu\right)^{\, 1/2}
\left(\frac{\alpha_\mathrm{ej}}{0.01}\right)\,\left(\frac{M_\mathrm{ej}}{0.1\,\Msun}\right) 
\nonumber \\
&\phantom{=}&
\phantom{211\,\mathrm{km\,s}^{-1}\,
\left(\frac{f_\mathrm{kin}}{\epsilon_5\,\beta_\nu}\right)^{\! 1/2}}
\!\!\!\!\!\!\!\!\,\,\,\,\,\times\,
\left(\frac{M_\mathrm{NS}}{1.5\,M_\odot}\right)^{-1} 
\label{eq:Janka}
\end{eqnarray}
where $\epsilon_5$, $f_\mathrm{kin}$, and $\beta_\nu$ denote the energy released per nucleon, the fraction of explosion energy that is kinetic energy, and the fraction of the ejecta mass that is neutrino heated, respectively. As per \citet{Janka2017}, we adopt constant values for $\epsilon_5 \sim 1$, $f_\mathrm{kin} \sim 0.1$ and $\beta_\nu \sim 0.1$. Following the discussions in \cite{Gessner&Janka2018} and \cite{Janka&Kresse2024}, we set the momentum asymmetry parameter $\alpha_\mathrm{ej}$ to 1\%, although we note that this parameter is poorly constrained. We obtain the ejected mass $M_\mathrm{ej}$ for each model by subtracting the total stellar mass at core carbon exhaustion from the remnant baryonic mass calculated from our collapse prescription \citep[e.g., the prescription from][]{Patton&Sukhbold2020}. 

\vspace{2pt}

(ii) Linear kicks: Using a combination of constraints from the sample of known DNSs and single-star pulsars, low kick velocities from USSN, and the DNS merger rate, \cite{Richards+2023} calculate the magnitude of natal kicks as a function of the SN ejecta mass:
\begin{equation}
    v_{\rm kick} = \alpha \left(\frac{M_{\rm ej}}{M_{\rm remnant}}\right) \, + \, \beta
\end{equation}
where $\alpha$ and $\beta$ are free parameters with best-fit values $\alpha = 115 \, \mathrm{km\,s}^{-1}$ and $\beta = 15 \, \mathrm{km\,s}^{-1}$, respectively. $M_{\rm remnant}$ represents the final remnant NS mass.

\begin{figure*}[t!]
\centering
\includegraphics[width=0.80\linewidth]{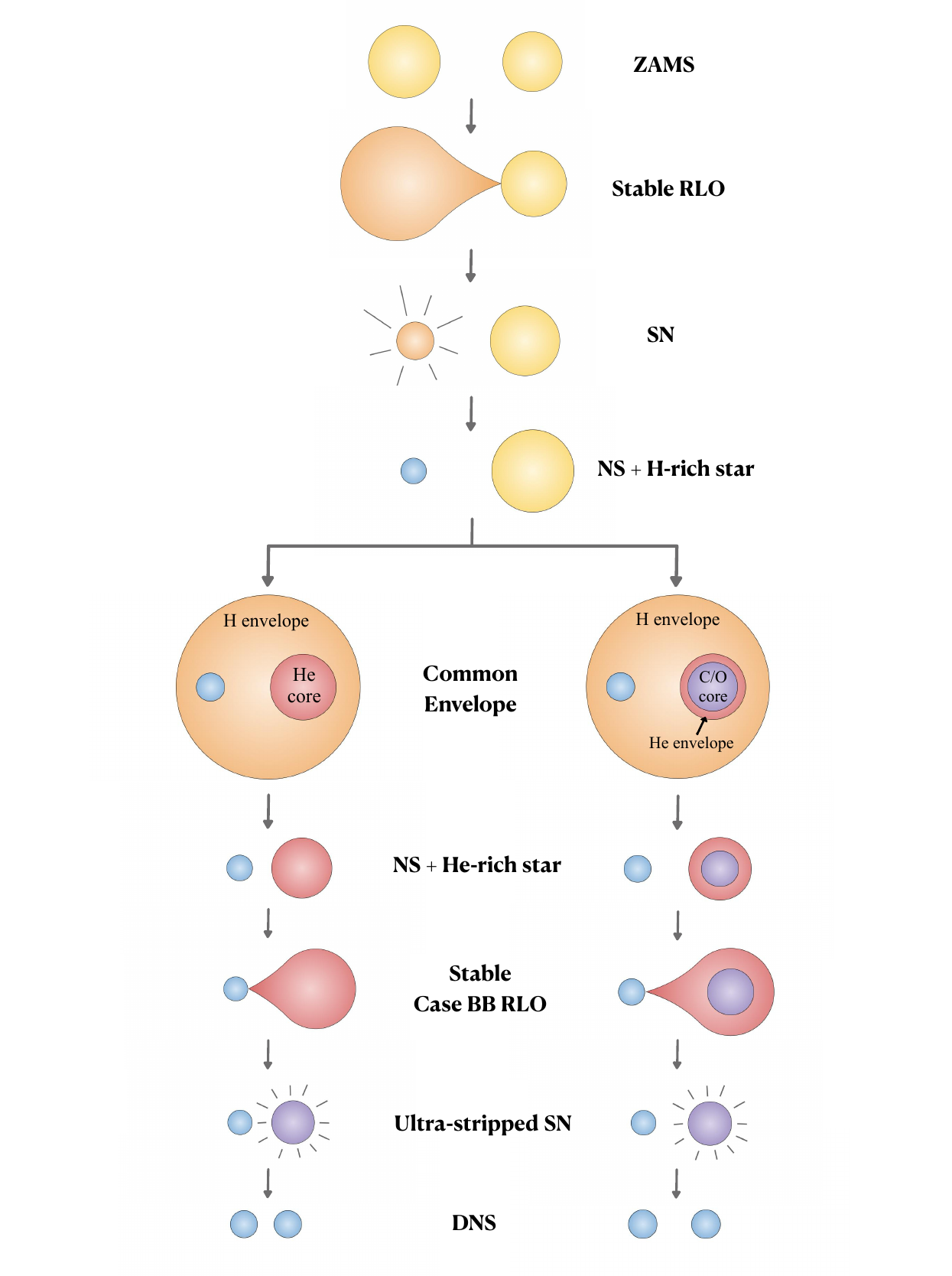}
\caption{Schematic diagram illustrating the complete evolutionary history starting with two hydrogen-rich stars at ZAMS leading to the formation of a DNS system within the two sub-channels in our model populations.   
The pivotal branching point occurs prior to the CE phase initiated by the hydrogen-rich star with the primary NS. Depending on the type of core the donor has developed, we classify the two channels as \Hecore channel (Case B RLO; left column) or \COcore (Case C RLO; right column). Following successful ejection of the common envelope, the primary NS is left in orbit with a helium star. We find that mass transfer at this stage remains stable in both channels leading to DNS formation.
With both the hydrogen and helium envelopes lost, the donor star now collapses in a stripped envelope supernova, leading to a DNS so long as the system is not disrupted. DNSs in the \COcore channel are born as a Type Ib SN, while those in the \Hecore channel can be observed as both Type Ib and Ic. As shown in Figure~\ref{fig:preferred_model}, the population of DNSs that merge within a Hubble time are produced via the \Hecore model, while DNSs formed through the \COcore model will not merge in the lifetime of the Universe.
}
\label{fig:schematic}
\end{figure*}

\subsection{Enhancements to our binary evolution grids}
\label{subsec:additional_grids}

To accurately model the entire population of DNSs, the \texttt{CO-HMS\_RLO} grid in \posydon Data Release 1 (DR1) consisting of a NS with a H-rich companion under-resolve a portion of parameter space. We therefore ran a supplemental grid of densely spaced models in the range $400 \lesssim P_{\rm orb}/\rm days \lesssim 3000$ and $5 \lesssim M_{\rm donor}/\Msun \lesssim 10$, with the NS companion mass ranging between $1.1 \lesssim M_{\rm NS}/\Msun \lesssim 2.0$, resulting in an additional 2250 models. These models were run with an identical set-up to that described in \citet{POSYDON+2023} and were combined with the \posydon DR1 grid to produce an enhanced grid with the resolution necessary for robust modeling of DNSs.\footnote{This dataset is available on Zenodo under an open-source Creative Commons Attribution license at doi: \href{https://doi.org/10.5281/zenodo.17240253}{10.5281/zenodo.17240253}.} We re-trained our interpolation schemes which were then used to generate the populations described throughout this work. 

Additionally, informed by improvements in the DR2 grids, we implemented the \texttt{dedt\_hepulse} rerun in our \texttt{CO-HeMS} grid consisting of a compact object and a He-rich star \citep[see][]{Andrews+2025} by adjusting the MLT++ conditions described in \cite{Paxton+2013}. This modification allows for the handling of superadiabatic temperature gradients for many stripped He star models and is particularly effective for resolving the set of non-converged \mesa models in the \texttt{CO-HeMS} grid. Finally, we note that this work includes a treatment of reverse mass transfer in the \texttt{HMS-HMS} grid of binaries composed of two H-rich stars as described in \citet{Xing+2024}, although contributions to DNS formation from such reverse mass transferring systems are negligible.

\subsection{Comparing our Model Populations to the Observed Sample}
\label{subsec:comparison_to_observed_sample}

A comprehensive, quantitative analysis comparing the 25 observed DNS systems with our model populations requires a Bayesian framework, which has been described in previous studies \citep{Andrews+2015, Alejandro+2018, Cecilia+2023, Deng+2024}. We forego that approach here as \posydon currently lacks a detailed pulsar model describing pulsar spin and magnetic field evolution, both of which affect their observability. Therefore, our principal results and conclusions are derived from a qualitative, visual comparison between the observed systems and our model populations. Nevertheless, as we demonstrate in Sections \ref{sec:ce} and \ref{sec:casebb}, even a qualitative comparison with the observed DNS distribution imposes stringent constraints on binary evolution. 

To guide our comparison, Figure~\ref{fig:preferred_model} provides both the distribution of observed systems and model systems in the $P_{\rm orb}-e$ space. We focus on these two variables as they are measured with high precision for every observed system. In the left panel we display the synthesized population for \texttt{MODEL12} (see Table \ref{tab:rates}) immediately after the second NS has formed (``after 2nd supernova"), while the right panel displays the systems as one might see them today (``present day''), accounting for our adopted constant star formation rate. Systems with short orbital periods quickly merge due to GW inspiral, so comparison between the two panels shows a precipitous drop in the number of overall systems that will merge within $\sim$10 Gyr (light gray markers). Although a similar drop ought to characterize the non-merging systems (dark gray markers), our lack of a pulsar model means we cannot account for systems that have spun down beyond the pulsar death line. Therefore, the relative ratio between the merging and non-merging modeled systems is not representative of the observational sample. Rather,
we focus on their general overlap with the observed DNSs in $P_{\rm orb}-e$ space.

\section{Evolution of DNS forming systems in our detailed binary star grids} \label{sec:grids}

\begin{figure*}[ht!]
\includegraphics[width=1.00\linewidth]{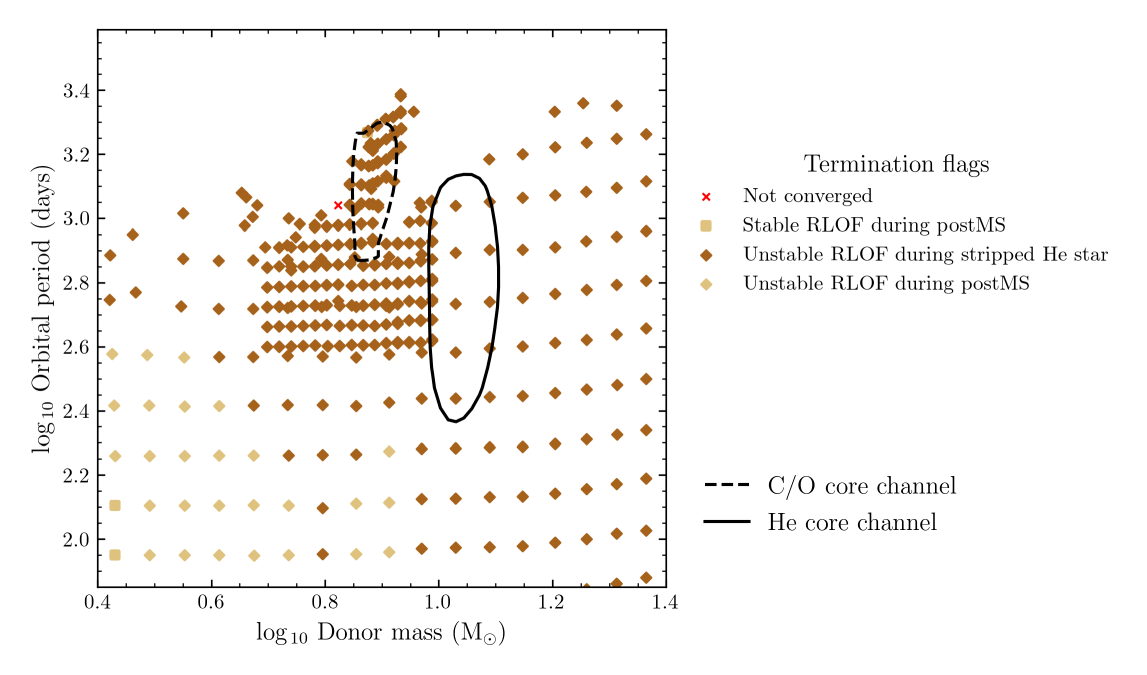}
\caption{A 2D slice summarizing the evolution of systems from our grid of binary star models consisting of a hydrogen-rich star and a $1.46\,\Msun$ NS at the onset of Roche lobe overflow. This is a composite grid made by combining the standard \posydon DR1 grid \citep[see Figure 11;][]{POSYDON+2023} and our denser set of models near the top (refer Section \ref{subsec:additional_grids} for details). The denser region of this grid has been plotted in detail in Fig. \ref{fig:CO-HMS_RLO_denser}. The contours denote the distribution of the donors from the two sub-channels in \texttt{MODEL12} at the $\mathrm{75th}$ percentile
derived from 2D histograms. Binaries that did not undergo any type of mass transfer are excluded from this plot. The labels denote the state of the donor at the last timestep in the simulation.
\label{fig:CO-HMS_RLO}}
\end{figure*}

For each of our model populations, we initialize and evolve $2 \times 10^6$ binaries, selecting the subset of the resulting population that produces DNSs. 
Table~\ref{tab:rates} summarizes the complete set of model populations along with their physical assumptions. Across all models, we find a consistent trend where DNSs predominantly form through a CE phase involving the first-born NS and its H-rich companion. While alternative formation channels do contribute, they play a comparatively minor role; we discuss these in more detail in Appendix \ref{sec:app_form_chan}. Guided by our model populations, we present below a qualitative description of our results on DNS formation with \posydon.

\subsection{Origin of DNS in our model populations}
\label{subsec:formation_channels}

While there exist multiple evolutionary pathways for DNS formation explored throughout the literature  \citep[e.g.,][]{Bhattacharya&vandenHeuvel1991}, we find that DNS formation in our models predominantly ($\gtrsim95\%$) follow one of two variations on the diagram shown in Figure~\ref{fig:schematic}. These binaries start as two main sequence stars on the ZAMS with masses typically $\sim$12$M_{\odot}$. They need to have an orbital period sufficiently small ($P_{\rm orb}\lesssim1500\ \mathrm{d}$; see discussion in Appendix~\ref{sec:HMS_HMS} and associated Figure~\ref{fig:HMS-HMS}) that the initially more massive star will overfill its Roche lobe upon expanding into a giant phase, but not so small that the system enters a contact phase (such binaries are assumed to merge due to the absence of a well-defined core). This mass transfer phase must be stable, so the initial mass ratio cannot be too far from unity (typically $q \gtrsim$ 0.7). 

After the primary star completes its evolution and collapses into a NS, \posydon\ directs the binary to the detached binary step, in which the binary is evolved semi-analytically, accounting for the effects of the secondary's evolution as well as their effect on the orbit (e.g., through tides and gravitational wave-driven orbital decay, although these effects are insignificant for the binaries forming DNSs in this phase). Upon expansion of the secondary as it evolves into a giant star, the binary enters a second phase of Roche lobe overflow, and \posydon transfers the binary into the CO-HMS grid (compact object with a H-rich star). 
Here, we find a bifurcation depending on whether the secondary initiates RLO while still having a helium core\footnote{Most of these systems are ascending the Red Giant branch, with a small fraction ($<3\%$) undergoing core helium burning.} or after forming a carbon-oxygen core. 
Throughout the text, we refer to these two formation pathways as the \Hecore channel (Case B RLO) and \COcore channel (Case C RLO), respectively. 
The difference is predominantly dictated by the orbital period, however, the resulting evolutionary outcomes diverge significantly. We discuss the specifics of this bifurcation further in Section \ref{sec:ce}. Regardless of the evolutionary state when the secondary ($\gtrsim$ 7.5$\Msun$) overfills its Roche lobe with a $\sim$ 1.4\Msun\, NS companion, the ensuing mass transfer phase is always unstable due to the large difference in component masses leading to a CE phase.

After \posydon evolves the binary through its common envelope routines in which the hydrogen envelope is lost, the binary is transferred to our grid of CO-HeMS binaries (compact object with a He-rich star). Upon expansion of the stripped He stars, these binaries enter a third phase of mass transfer, which is always stable (note, we are referring only to binaries that ultimately form DNSs. \posydon produces some binaries that enter unstable Case BB RLO, but these always merge; see Section~\ref{subsec:CO-HeMS}). Eventually these secondaries collapse in a supernova, leaving behind a DNS if the system remains bound.

\subsection{Our High Density Grid of NS Binaries with a H-Rich Companion} \label{subsec:CO-HMS_RLO}

\begin{figure}[h!]
\includegraphics[width=1.00\linewidth]{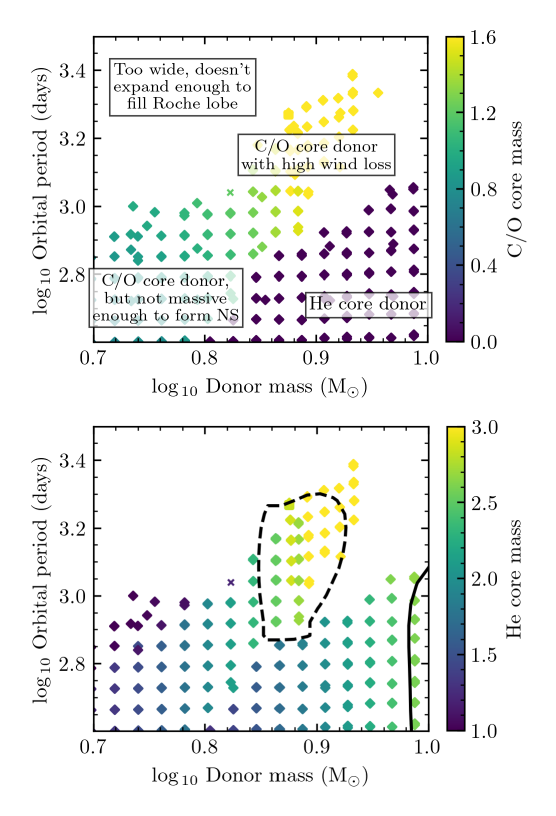}
\caption{A 2D grid slice showing the denser region of models from Fig.~\ref{fig:CO-HMS_RLO}, with similar marker legends and contours. The top panel displays the C/O core mass of the donor star at the onset of RLO. Text boxes in different regions of the plot offer explanations as to why the narrow strip of models protrudes from the rest, giving it a tail-like shape. In the bottom panel, we show the He core mass of the donor. Notably, in both panels, the tip of the tail is comprised of donors with massive cores which are likely to undergo iron core-collapse and produce massive compact objects. 
}
\label{fig:CO-HMS_RLO_denser}
\end{figure}

\begin{figure*}[ht!]
\includegraphics[width=1.00\linewidth]{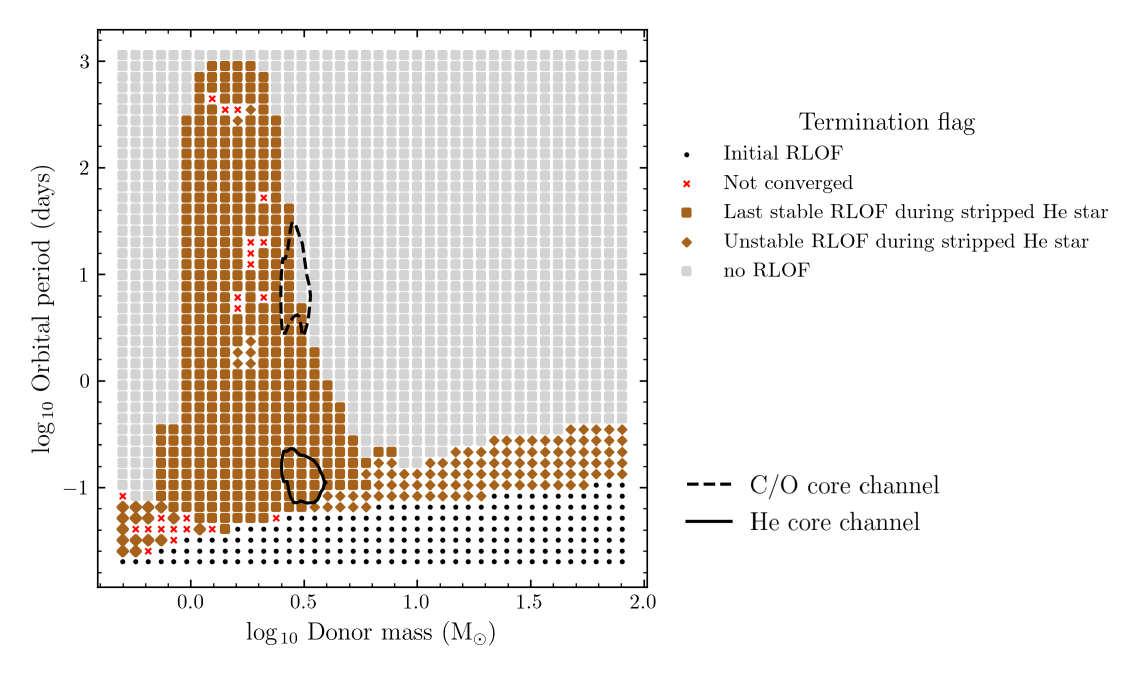}
\caption{A 2D grid slice from our binary star models consisting of a helium-rich star at zero-age helium main sequence (ZAHeMS) and a $1.46\,\Msun$ NS. The contours denote the regions enclosing $75\%$ of the donors at this evolutionary stage within the two sub-channels in \texttt{MODEL12}.
\label{fig:CO-HeMS}}
\end{figure*}

While we provide the details of how the DNSs in our models populate our initial grid of two H-rich stars at ZAMS in Appendix \ref{sec:HMS_HMS}, here we focus on the grid comprised of NSs with H-rich companions which includes our supplemental, high-density grid of binary models shown in Figure \ref{fig:CO-HMS_RLO}. 
The mass of the compact object in this figure is set to $1.46~\Msun$, reflecting the typical mass of the first-born NS at the onset of the MT episode. Since this grid presents the conditions right before RLO (following the approach in \posydon, any evolution in this grid prior to RLO is discarded), binaries that never underwent MT are excluded. Furthermore, the models are not exactly regularly spaced, as some mass loss due to winds makes the systems' masses and orbital periods at RLO somewhat different from their values at initialization.

The black lines in Figure~\ref{fig:CO-HMS_RLO} indicate the origin of our DNS populations for the two separate sub-channels displayed in Figure~\ref{fig:schematic}. In both scenarios, we find that all binary systems proceed via the formation of a CE (all markers are diamonds). 
However, compared to those in the \Hecore channel, the binaries in the \COcore channel (dashed line) tend to occupy a smaller region of the parameter space, corresponding to a distinct outcropping of models at high orbital periods. 

Their larger orbital period allows these stars sufficient time to develop a metal (C/O) core prior to RLO. In contrast, binaries from the \Hecore channel have less time to evolve before filling their Roche lobe and can only develop a helium core. We demonstrate this explicitly in Figure \ref{fig:CO-HMS_RLO_denser} where we present a detailed view of the denser set of models. The top and bottom panels use color coding based on the C/O core mass and helium core mass of the donor, respectively, with the legend labels and black lines retaining their meaning from Figure~\ref{fig:CO-HMS_RLO}.

The peculiar shape of the outcropping of models with developed C/O cores are the result of a number of separate effects. Binaries with larger orbital periods are too wide to fill their Roche lobes (white space in Figure \ref{fig:CO-HMS_RLO_denser}). On the other hand, those in the bottom-left corner develop a C/O core but fall below the mass range required for NS formation, ultimately becoming NS--WD systems. The models in the bottom-right corner overfill their Roche lobes while still having a helium core (with a median mass of $\sim 2.9\,\Msun$), constituting the \Hecore channel. 
In our \COcore binaries, the donor star is sufficiently distant during its post-MS expansion to experience prolonged mass loss from stellar winds prior to the onset RLO, reducing the stars' masses and widening their orbits. This leads to an upper-leftward shift away from the other binaries, causing these systems to protrude along the thin strip. These systems feature the most evolved donors, which are on the asymptotic giant branch (AGB) with a C/O core (median mass $\sim 1.4\,\Msun$).

\subsection{Case BB Mass Transfer \footnote{Although we refer to this phase as Case BB, some systems will in fact undergo Case CB mass transfer. We have adopted the Case BB terminology throughout this work as it is the commonly accepted convention in the field.}}
\label{subsec:CO-HeMS}

After expelling the CE, the NS is left in orbit around the former H-giant's exposed core. This star now evolves as a compact, hot, helium star. Upon core helium exhaustion, it expands its outer layers into a He-giant, and, for sufficiently close binaries, initiates Case BB RLO. Figure \ref{fig:CO-HeMS} shows models of helium-rich stars in orbit with a $1.46\,\Msun$ NS, meant to represent the median NS mass at this evolutionary stage in our model populations. Similar to previous figures, the regions of the parameter space leading to DNS formation are marked by black contours: dashed for the \COcore channel and solid for the \Hecore channel. The vast majority of systems engage in stable mass transfer, as indicated by brown squares. 

We find that an insignificant fraction ($<5\%$) of systems with massive helium stars and long orbital periods avoid mass transfer (gray squares), primarily because more massive helium stars tend to expand to smaller  radii, as illustrated in Figure 8 of \cite{POSYDON+2023}. It has been suggested that such massive helium stars may experience significant SN fallback accretion upon core collapse, and could possibly explain massive DNS that are also GW sources \citep{Alejandro+2021}.
While our \posydon model populations show that they do yield a more massive secondary NS ($\sim 1.41\,\Msun$ compared to $\sim 1.26\,\Msun$ for others), we track the further evolution of these systems, and find that in our models they form DNSs with orbits too wide to merge in a Hubble time.

This phase of Case BB mass transfer is crucial in bringing the binary closer to form short period progenitor systems which lead to the formation of merging DNSs. Previous works have proposed unstable Case BB mass transfer as a pathway to produce fast-merging DNSs, which can explain the formation of LIGO systems like GW190425 \citep{Ivanova+2003, IRS+2020, Galaudage+2021}, or the $r$-process enrichment of ultra-faint dwarf galaxies \citep{Safarzadeh+2019}. In agreement with previous detailed studies of the Case BB mass transfer phase \citep[e.g.,][]{Tauris+2015} our detailed binary evolution grids of He-rich stars with NS companions demonstrate that such an unstable Case BB RLO will either end in a merger inside the CE, or leave behind an ultra-stripped core that is not massive enough to form a NS (forming WDs instead, making this a potential pathway for short-period NS--WD binaries). Our results therefore suggest that unstable Case BB mass transfer cannot be responsible for forming fast-merging DNSs.

\section{Constraining common envelope evolution} \label{sec:ce}

\begin{figure*}[ht!]
\includegraphics[width=1.00\linewidth]{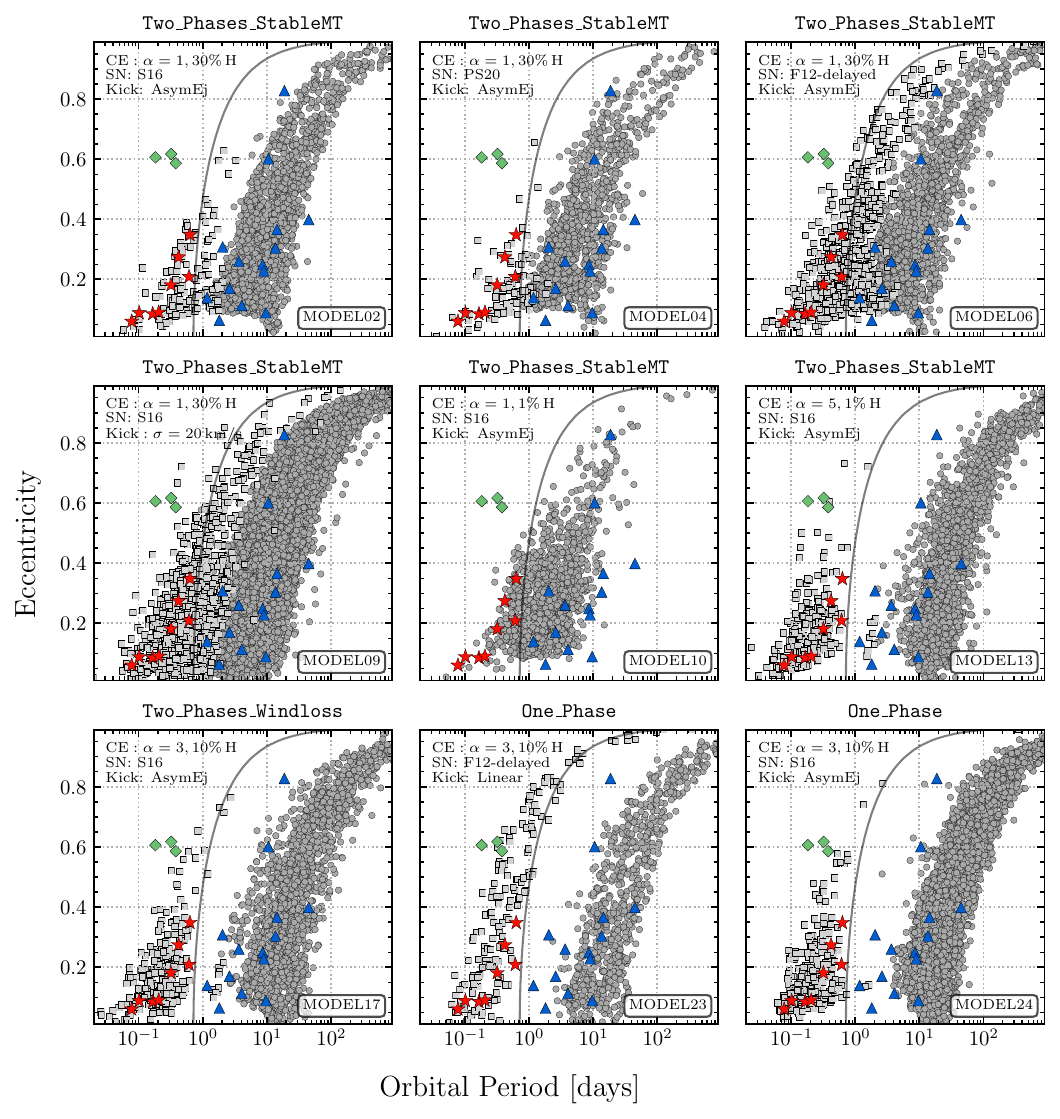}
\caption{The present day distribution of DNSs from selected models on the $P_{\rm orb} - e$ plane, with marker legends following the same convention as in Figure \ref{fig:preferred_model}. The gray line in each panel indicates a merger time equal to $t_{\rm Hubble}$ calculated for $1.4\Msun$ NS components. For a summary of the models and their physical assumptions, refer to Table \ref{tab:rates}.
The plots demonstrate the bifurcation of the DNS distribution into the \Hecore (light gray markers) and \COcore (dark gray markers) subchannels, sometimes with a distinct orbital period gap. Since the distributions shown correspond to the present day DNS population; many systems from the \Hecore channel may have merged since formation and are thus missing from these plots. Nevertheless, a consistent trend across models is the separation of the two formation channels; the \Hecore systems tend to survive at shorter orbital periods, occupying regions mostly near sub-population (i) of the observed sample (red star markers), whereas the \COcore systems typically have longer periods associated with sub-population (ii) (blue triangle markers). This bifurcation is consistent across varying treatments of the common envelope, natal kick distributions, and core-collapse prescriptions. Notably, in every model exhibiting this bifurcation, the merging DNS systems are formed exclusively through the \Hecore channel.}
\label{fig:all_models}
\end{figure*}

\subsection{General Trends} \label{subsec:trends}

Even without the clustering analysis from \citet{Andrews&Mandel2019}, the diversity of the observed systems in Figure \ref{fig:Obs_Porb-e} hints that multiple, distinct evolutionary channels are likely required to fit the entire population of DNSs. Specifically, sub-populations (i) and (ii) when combined, span an orbital period range of nearly three orders-of-magnitude. Indeed, previous population synthesis studies have identified several different evolutionary channels that could form DNSs, including evolution through a CE \citep[e.g.,][]{Bhattacharya&vandenHeuvel1991}, stable mass transfer \citep[e.g.,][]{Andrews+2015}, a double-core CE \citep[e.g.,][]{Alejandro+2018}, or no mass transfer at all \citep[e.g.,][]{Stevenson+2022}.

Comparison between our synthetic populations and the observed systems in Figure~\ref{fig:preferred_model} demonstrates a broadly consistent trend where systems with the shortest orbital periods and highest eccentricities merge quickly due to general relativistic circularization and orbital decay, while supernova kicks tend to produce a greater spread in eccentricity for systems with longer orbital periods \citep{Andrews&Zezas2019}. Sub-population (iii), with a tightly clustered sample of three systems in orbital period-eccentricity space, bucks this trend; none of the models we tested can form systems matching their parameters in statistically significant numbers. We therefore reinforce the conclusion from \citet{Andrews&Mandel2019} that the three systems comprising sub-population (iii) are a challenge to produce through isolated binary evolution. However, we note that Doppler smearing leads to strong observational biases at these orbital periods and eccentricities \citep[][]{Bagchi+2013}; a complete study would incorporate these selection effects along with a pulsar evolution model \citep[for an example of such a model applied to NS-BH systems, see][]{Chattopadhyay+2021}. Barring a complete accounting of selection effects, throughout the remainder of this work we focus on the formation of sub-populations (i) and (ii), leaving the difficulty of explaining sub-population (iii) for future work.

\subsection{Details of the bifurcation} \label{subsec:bifurcation}

As described in Section~\ref{subsec:formation_channels}, we find that at Solar metallicity, DNSs are predominantly formed through a common envelope scenario, with the \Hecore and \COcore sub-channels being determined by the evolutionary state of the donor at the onset of RLO.
Figure \ref{fig:all_models} shows the present day distribution of orbital periods and eccentricities from selected model populations exploring different treatments of the CE, natal kick distributions, and SN core-collapse modeling. 
All model populations consistently produce systems through the \COcore channel (dark gray markers), which represents the standard formation pathway for DNSs involving mass transfer from an asymptotic giant branch (AGB) donor. However, some models such as \texttt{MODEL 10} shown in Figure~\ref{fig:all_models} do not form systems through the \Hecore channel (light gray markers). While these models cover a wide range of orbital periods, they cannot cover the entire range of observed systems, missing systems at one or both extremes. Consideration of the other models in Figure~\ref{fig:all_models} shows that no one subchannel can cover the entire range of systems. 

We expect the inability of any individual subchannel to form the entire range of systems to be a robust result of binary evolution theory: the range of DNS orbital periods is directly dependent upon the range of post-CE orbital periods, which itself is determined by the range of envelope binding energies for CE donors. These binding energies naturally bifurcate depending on whether the donor star has either a He core and therefore a relatively more tightly bound envelope leading to the short-period systems in Figure~\ref{fig:all_models}, or a C/O core with a less bound envelope, leading to the formation of wider DNSs. While we conclude that one subchannel cannot individually span the range of the observed systems, perusal of Figure~\ref{fig:all_models} shows that when combined, the two subchannels can explain both the shortest and widest period systems.

While we find that the \Hecore channel emerges across most CE treatments in Figure~\ref{fig:all_models} and a range of $\mathrm{\alpha_{CE}}$ values, its formation is favored by models that promote efficient CE ejection (see discussion in Section \ref{subsec:binding_energy}). This condition can be achieved in one of two ways: either $\mathrm{\alpha_{CE}}$ is set to be larger than unity, or we adopt a generous definition for a star's core. Based on the results of \citet{Fragos+2019}, our default model opts for the latter---a two-phase CE prescription with a 30\% H fraction determining the core-envelope boundary. If we alternatively adopt a 1\% or 10\% H fraction as the core-envelope boundary (see \texttt{MODELS 13, 17, 23} and \texttt{24} in Figure \ref{fig:all_models}), then a higher $\mathrm{\alpha_{CE}} \gtrsim1.2$ is required. 

In models without one of these conditions, systems that would have otherwise populated the \Hecore channel instead merge inside a CE. 
We do not attempt to fine-tune the value of $\mathrm{\alpha_{CE}}$ to quantitatively match the observed distribution, recognizing instead that $\mathrm{\alpha_{CE}}$ serves as an ignorance parameter encompassing complex physical processes. We only point out that the broad characteristics of sub-populations (i) and (ii) are largely matched by the \Hecore and \COcore channels, and the systems forming through the \COcore channel may experience a somewhat different efficiency than those forming through the \Hecore channel. We leave the more difficult problem of model selection, which involves a careful inclusion of observational biases, to future work. Here, we conclude from our models that efficient CE ejection achieved through a number of ways of treating the CE phase leads to the emergence of the \Hecore channel. 

Finally, we note that since a more efficient CE ejection model shifts the entire distribution to larger orbital periods leading to wider separations and fewer systems merging during the CE phase, there is a strong dependence of the efficiency on the rate of DNS formation. Therefore, in principle, the overall DNS merger rate and delay time distribution can constrain the CE model, particularly for the \Hecore channel which dominates the population of merging DNSs. However, additional constraints are needed to break the degeneracy between the efficiency parameter $\alpha_{\rm CE}$ and the core-envelope boundary definition. 

\begin{figure*}[ht!]
\includegraphics[width=1.0\linewidth]{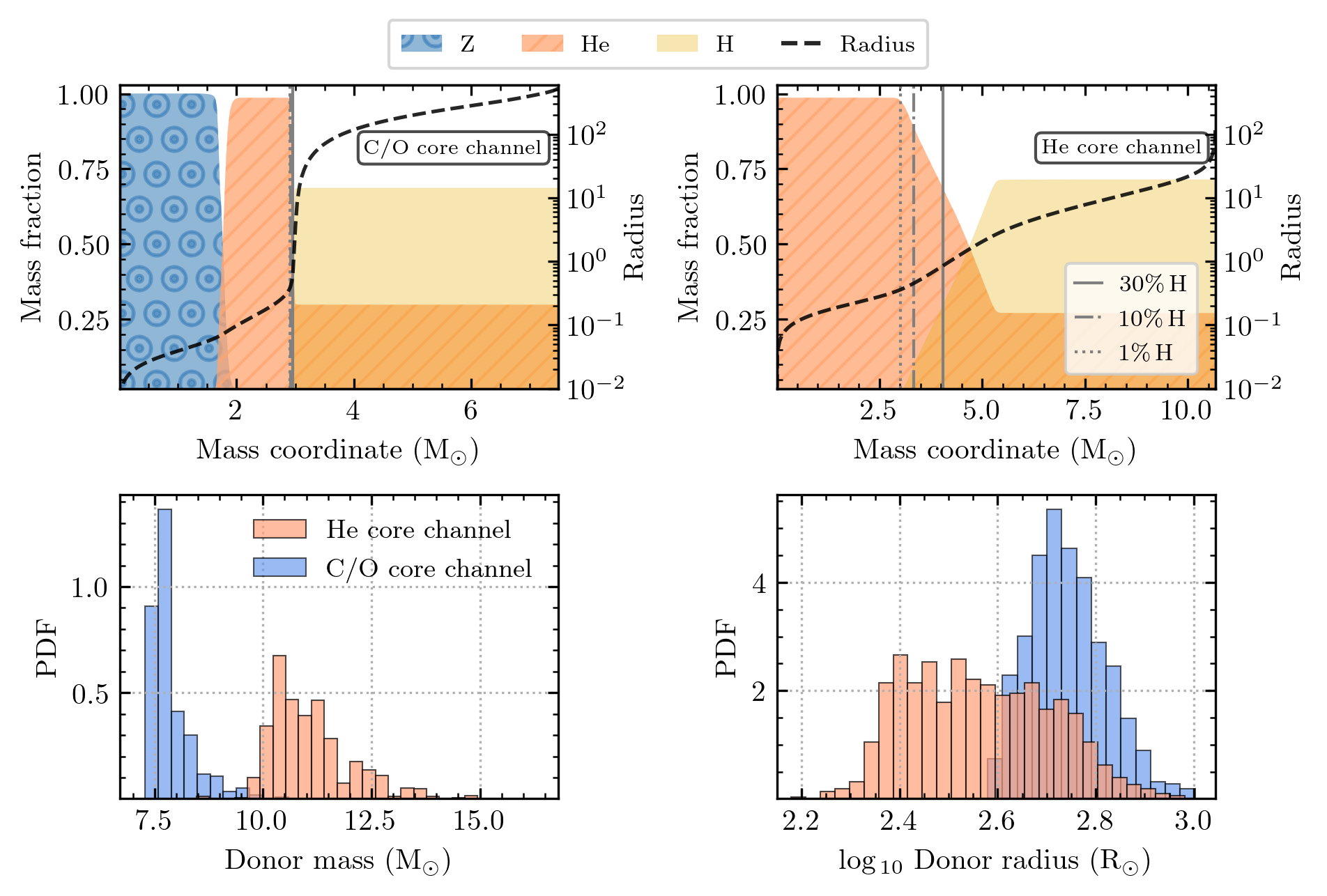}
\caption{(\textit{Top row}) Stellar composition of typical companions to the first-born NS at the onset of mass transfer in the two sub-channels are shown. The bifurcation in the final DNS orbital periods is characterized by the evolutionary stage of the donor at the onset of RLO; while some donors have helium depleted, metal-rich cores (left), others initiate mass transfer early in their evolutionary sequence with a core rich in helium (right). The three vertical lines depict the core-envelope boundaries in our models, defined at $1\%$, $10\%$ and $30\%$ H-fraction. (\textit{Bottom row}) Distributions of the donor masses and radii at the onset of RLO for the two sub-channels in \texttt{MODEL12}. The typical donor in the \COcore channel is less massive and has expanded to a larger radius than those in the \Hecore channel, having underwent prolonged expansion and wind-driven mass loss.}
\label{fig:oRLO2}
\end{figure*}

\subsubsection{Sub-channel with mass transfer from a donor with a C/O core} \label{subsubsec:co_core_channel}

The top and bottom rows of Figure \ref{fig:oRLO2} illustrate the properties of the two typical types of donor stars at the onset of mass transfer onto the primary NS. The left panel of the top row shows the structure of a donor that has developed a $\sim 2\Msun$ C/O core, with a $\sim 1\Msun$ He shell, surrounded by a $\sim 4.5\Msun$ H-envelope, prior to RLO. This type of donor entering a CE with the primary NS represents the well-known path to DNS formation involving unstable mass transfer with an AGB star (as depicted in the top panel of Figure \ref{fig:CO-HMS_RLO_denser}). After the CE phase, the H-envelope is removed at the expense of orbital energy, leaving behind a C/O core with a naked helium envelope in a tight orbit with the primary NS. In the $P_{\rm orb} - e$ diagrams, these are the systems marked with dark gray markers (Figures \ref{fig:preferred_model} and \ref{fig:all_models}). 

Following the birth of the first NS, the binary separation is large enough for the donor star to develop a C/O core before RLO. During this long, detached phase (lasting $\mathrm{\sim 9~Myr}$), the donor undergoes significant expansion into its AGB stage, and is expected to lose a substantial amount of mass through stellar winds before initiating RLO. The bottom row of panels in Figure \ref{fig:oRLO2} displays the distribution of the masses and radii of the donor stars immediately before the onset of RLO from \texttt{MODEL12}. We find that donors with a developed C/O core are less massive (left panel) and have expanded to larger radii (right panel). Additionally, the donor spends around $\mathrm{\sim 3000 \, yr}$ between the initiation of RLO and the onset of the CE. Since in this work we assume no CE accretion (although this assumption might be too simplistic, see discussion in Section \ref{subsec:POSYDON}), the period of stable mass transfer leading up to the CE phase is the only time when the NS accretes significant mass from the H-rich star (we further argue in Appendix \ref{sec:appendix_wind_accretion} that wind accretion is negligible). 

\subsubsection{Sub-channel with mass transfer from a donor with a He core} \label{subsubsec:he_core_channel}

The right panel of the top row in Figure \ref{fig:oRLO2} shows the second kind of donor entering into a CE with the primary NS; a younger star with a $\sim 3 \Msun$ helium core inside a $\sim 7.5 \Msun$ envelope of mainly hydrogen. Before engaging in mass transfer, the system remains detached for $\mathrm{\sim 3~Myr}$, roughly one-third of the time seen in the \COcore channel. 
The comparatively short orbital periods (see Figure \ref{fig:CO-HMS_RLO}) mean the donor evolves only enough to develop a helium core before filling its Roche lobe.
It loses comparatively less mass to stellar winds and does not expand its outer layers as much as an AGB star. The bottom row of Figure \ref{fig:oRLO2} reveals that in our model populations, these donors are typically more massive (left panel) and smaller in radius (right panel) compared to those in the \COcore channel. At the onset of RLO, the donor is typically in the Red Giant Branch (RGB).
Furthermore, once RLO commences, mass transfer quickly becomes unstable, within $\mathrm{\sim 300\,yr}$, about one-tenth of the time in the \COcore channel. 

\begin{figure*}[ht!]
\centering
\includegraphics[width=1.00\linewidth]{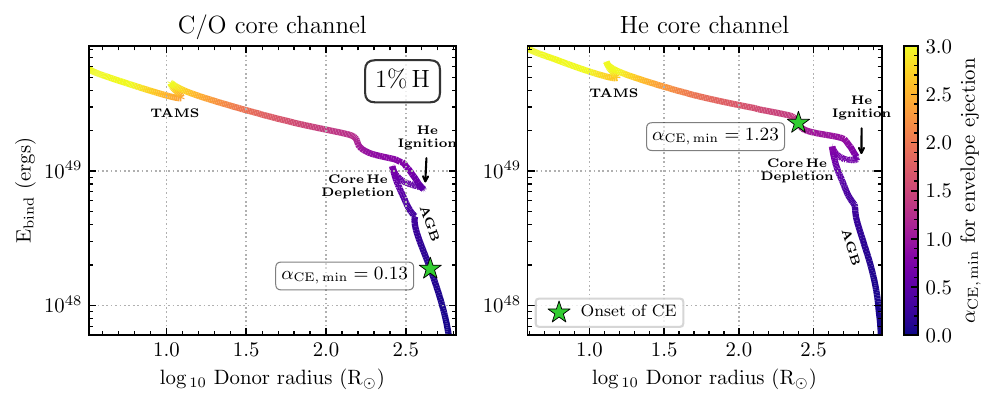}
\includegraphics[width=1.00\linewidth]{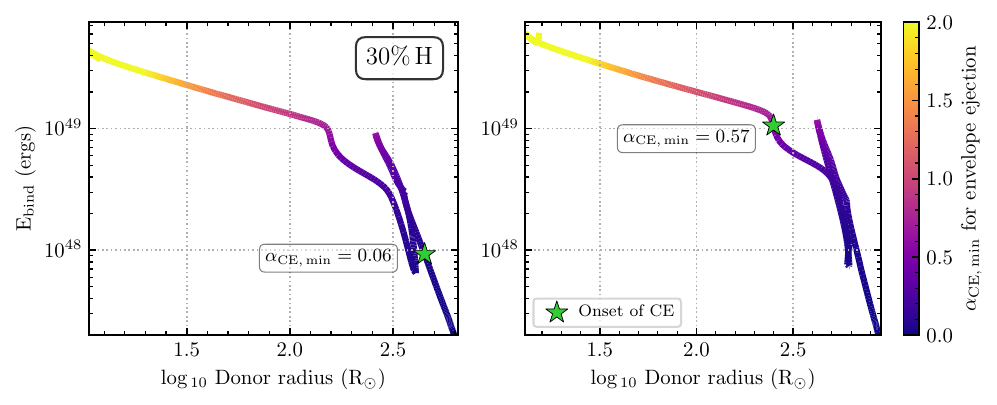}
\caption{(\textit{Top row}) The left and right panels follow the evolution of a single hydrogen-rich star with a mass typical of the companion of the first-born NS in the \COcore ($M \sim 8\,\Msun$) and \Hecore ($M \sim 11\,\Msun$) channel, respectively. The envelope binding energies of both stars decrease over time as the stars expand. Line color indicates the minimum value of $\alpha_{\rm CE}$ required to successfully eject the envelope, were that binary to initiate RLO at that timestep. Envelope binding energies are calculated assuming a 1\% H-fraction definition for the core, and $\alpha_{\rm CE}$ is calculated assuming a traditional $\alpha$--$\lambda$ prescription. Green stars indicate the median onset of RLO for each channel, leading to a common envelope. Because the donor in the \COcore channel overfills its Roche lobe in a more-evolved state, the system will survive a CE even for extremely low values of $\alpha_{\rm CE}$. The \Hecore channel, with its relatively less-evolved structure, requires a somewhat larger value of $\alpha_{\rm CE}\gtrsim1.2$ to survive.(\textit{Bottom row}) Same as the top row, but for a core-envelope boundary defined by a $30\%$ H-fraction. With this definition, the minimum value of $\alpha_{\rm CE}$ required to unbind the envelope for a donor in the \Hecore channel is $<1$. 
}
\label{fig:Ebind}
\end{figure*}

\subsection{Explanation using the envelope binding energies of the donors} \label{subsec:binding_energy}

\begin{figure*}[]
\includegraphics[width=1.00\linewidth]{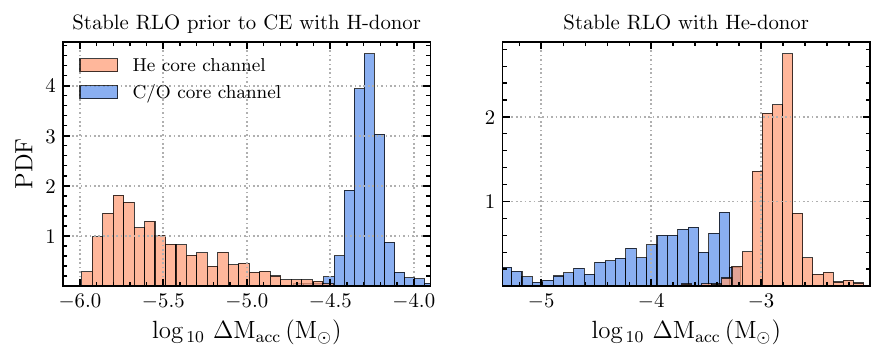}
\caption{Mass accreted by the primary NS in the two sub-channels during stable Roche lobe overflow from a hydrogen-rich donor (left) and subsequently from the helium star (right). The notable difference in magnitudes between the two panels suggests that pulsar recycling in DNS systems predominantly occurs after the companion has been stripped of its hydrogen envelope and initiates mass transfer as a helium star. We find that the maximum mass accreted during this stage can be up to 0.2$\,\Msun$.}
\label{fig:mass_accreted}
\end{figure*}

The disappearance of the \Hecore channel for many of our models in Figure \ref{fig:all_models} levies a limit on the CE ejection efficiency, which we find originates from the structure of the donor's envelope; the larger binding energy of the donor's envelope in the \Hecore channel requires a higher energy to expel the CE. In the top row of Figure \ref{fig:Ebind}, we show the evolution of the envelope binding energy calculated adopting a 1\% H-fraction definition for the core of a single H-rich star representative of the donor stars from the \COcore channel (8 $\Msun$; left panel) and \Hecore channel (11 $\Msun$; right panel).  As a star evolves (from left to right in these diagrams), its envelope becomes less bound. Assuming a $1.4\Msun$ NS companion, we calculate the maximum energy available to a binary on the verge of overfilling its Roche lobe, at every timestep. Comparing the resulting energy with the envelope binding energy at that timestep gives us the minimum $\mathrm{\alpha_{\rm CE}}$ (indicated by the line color) that would allow the post-CE binary product to survive. To calculate this minimum $\mathrm{\alpha_{\rm CE}}$, we have adopted a traditional one-phase CE using the $\alpha$--$\lambda$ prescription. We note that the exact values of $\mathrm{\alpha_{\rm CE,\,min}}$ shown in Figure~\ref{fig:Ebind} are strongly dependent on the core definition.

The hook on the right of each panel indicates core helium ignition. The green star marker denotes the radius at which a representative donor from each channel is expected to initiate a CE. Stars overfilling their Roche lobes after helium ignition have developed a C/O core (e.g., left panel) and have loosely bound envelopes that are easily ejected, even with $\alpha_{\rm CE}$ as low as $\simeq0.2$. The right panel, on the other hand, shows an example binary that overfills its Roche lobe prior to core helium ignition (from our \Hecore channel). These binaries have much higher envelope binding energies and require $\alpha_{\rm CE} \gtrsim 1.2$ to survive. However, we again emphasize that this $\alpha_{\rm CE}$ is calculated adopting a 1\% H-fraction definition for the core mass, and a traditional one-phase CE. A similar plot with a 30\% H-fraction definition is presented in the bottom row of Figure \ref{fig:Ebind}, which shows that under this definition, the envelope of the donor in the \Hecore channel can be unbound with $\alpha_{\rm CE}=1$. In fact, when adopting the default \posydon definition of a 30\% H fraction and a two-phase CE, the \Hecore channel is populated with $\alpha_{\rm CE}=1$ (\texttt{see MODELS 02, 04, 06} and \texttt{09} in Figure \ref{fig:all_models}).

\section{Mass accretion and pulsar recycling} \label{sec:casebb}

So far, our focus has been on the orbital parameters; however, the spin period and its derivative provide further clues about the origins of DNS systems. \posydon self-consistently models the mass accretion in binaries, allowing us to estimate the spin period of the first-born NS after undergoing multiple episodes of mass transfer. In our model populations, the primary NS undergoes two phases of mass accretion throughout its evolution: (i) a brief period of stable MT from the H-rich companion before the system becomes dynamically unstable, and (ii) mass transfer from the He star companion, or Case BB RLO \citep{DeGreve&DeLoore1977, Delgado&Thomas1981, Dewi+2002, Ivanova+2003, Tauris+2013}. As of now, \posydon assumes no mass is transferred during the CE phase itself \citep[which could be as large as $0.1\Msun$;][]{Macleod&Ramirez-Ruiz2015}. Additionally, for our analysis, any wind accretion onto the NS during the X-ray binary phases can be safely ignored (as demonstrated in Appendix~\ref{sec:appendix_wind_accretion}). 

Figure \ref{fig:mass_accreted} illustrates the amount of mass accreted by the primary NS from each of these two evolutionary phases: from the hydrogen-rich donor (left panel) and helium-rich donor (right panel). As discussed in Section \ref{subsec:CO-HeMS}, the Case BB RLO phase with the helium star donor is stable and proceeds on a thermal timescale ($\sim 10^4$ yr). This is much longer than the time it takes for MT to become unstable prior to the CE with a H-rich donor ($10^2 - 10^3$ yrs), which typically is of the order of the orbital decay timescale. As a result, most of the total mass accreted by the first-born NS occurs during Case BB RLO (under the assumption of insignificant mass accretion during a CE). Thus, mass transfer from the He-star is crucial for spinning-up the recycled pulsar which carries important observational consequences. 

In the \COcore channel, the NS--He star binary starts with a much wider orbit than in the \Hecore channel (see Figure \ref{fig:CO-HeMS}). This wider orbit allows the helium star more time to evolve before filling its Roche lobe, resulting in less time for mass transfer onto the NS before the He star terminates its nuclear burning and explodes in a supernova. As a result, the amount of mass accreted by the recycled pulsar from the He donor is generally lower in the \COcore channel (right panel of Figure \ref{fig:mass_accreted}), possibly leading to pulsars that form through the \Hecore channel experiencing more recycling.
This result is broadly consistent with the observations (see Figure \ref{fig:Obs_P-Pdot}); if sub-population (i) originates from the \Hecore channel, its shorter spin periods are an expected outcome.

\section{Supernova types \& natal kicks} \label{sec:supernova}
\begin{figure*}[ht!]
\includegraphics[width=1.0\linewidth]{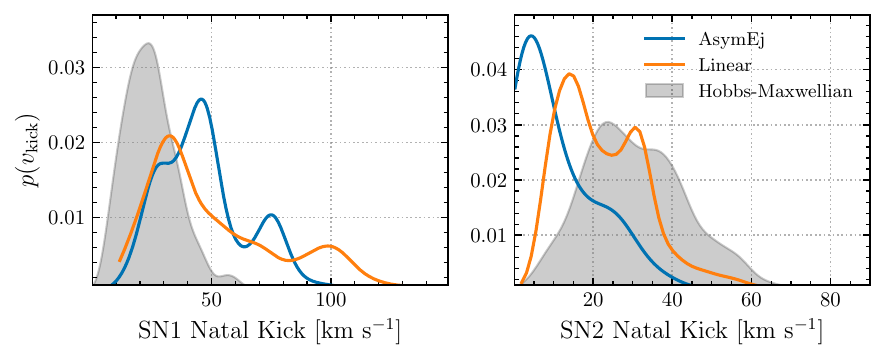}
\caption{Distribution of natal kicks for the first (left) and second supernova (right), respectively, in systems leading to DNS formation from the different natal kick prescriptions used in this work. A few data points extending beyond the depicted range in both rows were disregarded due to their relatively low numbers.
\label{fig:natal_kicks}}
\end{figure*}

Multiple studies have highlighted the critical role of the supernova stage in the dynamical evolution of DNSs \citep[e.g.,][]{Brandt+1995, Kalogera1996, Podsiadlowski+2004, Beniamini&Piran2016, Tauris+2017, Andrews&Zezas2019}. It therefore comes as no surprise that our resulting population is substantially altered by variations to the kick prescription as compared with the standard Hobbs-Maxwellian kicks commonly used in population synthesis codes. These variations include differences in supernova types, the fraction of USSN, merger rates, and other key factors. Given the pivotal impact of accurately modeling natal kicks, we provide a detailed analysis of how these kick models influence our population synthesis results and how they have potential to constrain DNS formation and evolution.

\subsection{On natal kicks} \label{subsec:natal_kicks}

As described in Section~\ref{subsec:POSYDON}, we test two ``physics-based" kick models in addition to the traditional isotropic, Maxwellian kick prescription with $\sigma_\mathrm{CCSN}=265\,\mathrm{km\,s^{-1}}$ \citep{Hobbs+2005}\footnote{However, note that recent work by \cite{Disberg&Mandel2025} has reported that the original \citet{Hobbs+2005} analysis omits a necessary Jacobian term, leading to a systematic overestimate of the inferred velocities.} for core-collapse SN, and $\sigma_\mathrm{ECSN}=20\,\mathrm{km\,s^{-1}}$ for NSs formed through ECSN. However, in agreement with \citet{Bray&Eldridge2016} we find this model fails to accurately represent DNS natal kicks.  As has been previously argued \citep{Tauris+2017}, to avoid merger within a CE, the system must have formed in a wide orbit, making it susceptible to disruption during the first SN by kicks of $\mathcal{O}(100)$ km s$^{-1}$. Indeed, while many stars collapse into NS through CCSN in our models with the Hobbs-Maxwellian distribution as the kick prescription, many of these systems tend to disrupt during the first SN, having received larger kicks. This is shown explicitly in the left panels of Figure~\ref{fig:natal_kicks}, where the range of kick velocities for DNS pairs is $0 - 60 \, \rm km/s$ for the Hobbs-Maxwellian prescription; any higher kick tends to disrupt the system. This conclusion is further supported by Table \ref{tab:rates} where we provide the fractional formation rate through CCSN for both the first-born (SN1) and second-born (SN2) NSs -- the models with the Hobbs-Maxwellian kick prescription tend to have higher ECSN (lower CCSN) fractions.

\begin{figure*}[t!]
\includegraphics[width=1.0\linewidth]{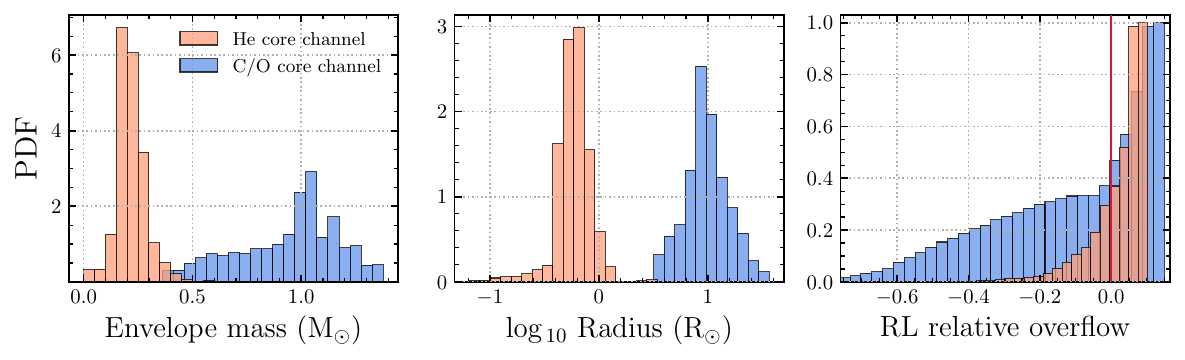}
\caption{Distribution of the helium envelope masses (left), stellar radii (center) and Roche lobe filling fraction (right) at the onset of the second supernova in the two sub-channels from \texttt{MODEL12}. The progenitor in the \Hecore channel tends to have less massive envelopes and are pretty compact as indicated by their radius distributions, hinting at extensive stripping of their helium envelopes prior to the second supernova. The vertical red line on the cumulative distribution plot on the right demarcates the progenitors with underfilled ($<0$) and overfilled ($>0$) Roche lobes. This highlights that $\sim 60\%$ of all progenitors were undergoing mass transfer at the time of the second supernova.}
\label{fig:SN_progenitor}
\end{figure*}

Several studies have suggested that the NS natal kicks are not only correlated with the explosion energy \citep{Janka2017, Muller+2019, Burrows+2023}, but also with the amount of ejecta mass lost during the SN \citep{Bray&Eldridge2016, Bray&Eldridge2018, Giacobbo&Mapelli2020, Mandel&Muller2020, Richards+2023}.
This motivates our implementation of the ``physics-based kick prescriptions'' in our populations. These prescriptions are agnostic as to whether a NS was formed through an ECSN or a CCSN; the same treatment is applied to all NSs. As a result, a clear difference in kick magnitudes between the primary and secondary SN becomes apparent (note the different dynamic ranges in the panels of Figure~\ref{fig:natal_kicks}).  
For instance, in the \textit{AsymEj} kick prescription, the natal kicks imparted are in the range of $10 - 90 \, \rm km/s$ for the primary SN and $0 - 40 \, \rm km/s$ for the secondary SN, naturally resulting in a bimodal kick distribution between SN1 and SN2. This difference is a result of assumptions built into the ``physics-based'' prescriptions where the kick velocity scales with the ejecta mass $M_{\rm ej}$ in the \textit{AsymEj} model, and with $M_{\rm ej}/M_{\rm remnant}$ in the \textit{Linear kick} model.
Having formed from a stripped star with a relatively small envelope, the second-born NS is given a significantly smaller natal kick. The impact of the kick velocity on DNS eccentricities and orbital periods has been well-studied elsewhere, both for individual systems \citep{Wex+2000, Willems&Kalogera2004, Willems+2004, Dewi&VanDenHeuvel2004, Wong+2010}, as well as for the population as a whole \citep[e.g.,][]{Andrews+2015, Beniamini&Piran2016, Alejandro+2018, Kruckow+2018, Shao&Li2018, Andrews&Zezas2019}. Here we only comment that while some individual systems are constrained to form through large kicks, the observed population as a whole is more consistent with kick velocities $\lesssim 50 \, \rm km/s$ \citep[see also][]{Disberg+2024, Guo+2024}. This result is consistent with findings from long-term 3D simulations study by \cite{Burrows+2023}, which supports a correlation between lower kick velocity and lower NS gravitational mass.

Additionally, we find that the ``physics-based kick prescriptions'' allow for NS formation through low-ejecta mass Fe core-collapse, reducing the contribution from ECSN. While our models with the Hobbs-Maxwellian kicks yield higher ECSN fractions for SN1 (see Table \ref{tab:rates}; in agreement with \citealt{Giacobbo&Mapelli2019} who report that the first SN is typically an ECSN), we attribute this to the strong Hobbs-Maxwellian kicks for Fe core-collapse, which tend to disrupt binaries during the first SN. However, this is no longer the case for models with ``physics-based kick prescriptions'', which instead show that the first SN is mostly Fe core-collapse.  For the second-born NSs, our \mesa binary simulations indicate that their progenitors more frequently fall within the ECSN mass range, leading to a higher ECSN contribution to the second SN compared with the first.
While \cite{{Tauris+2017}} show that DNS progenitors can typically survive even large kicks during the second SN; nonetheless, we find that the majority of DNSs are born with small natal kicks (right-hand panel in Figure~\ref{fig:natal_kicks}) predominantly originating from CCSN involving stripped progenitors with low ejecta mass, with some contribution from small ECSN kicks.

Since SN disruption strongly affects the DNS merger rate, we can independently check the consistency of our models by comparing the derived merger rates (provided in Table~\ref{tab:rates}) with the observed constraint from the LIGO-Virgo-Kagra Collaboration. We discuss this further in Section~\ref{sec:merger_rates}.

\begin{figure*}[ht!]
\includegraphics[width=1.0\textwidth]{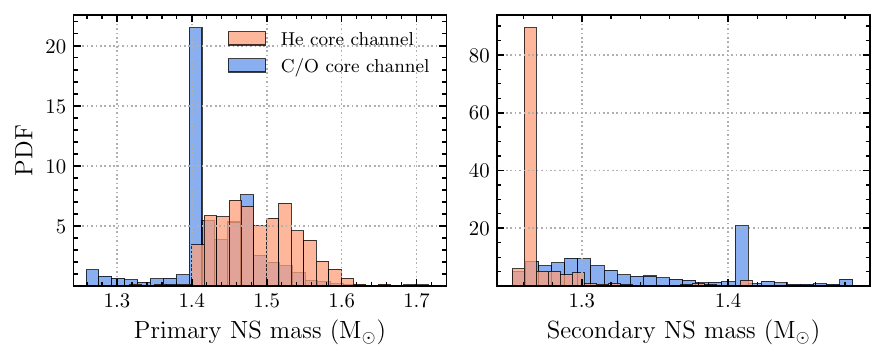}
\caption{The mass distributions of the two NSs in the \Hecore and \COcore sub-channels for  \texttt{MODEL12} (adopting the \citealt{Sukhbold+2016} prescription for core-collapse and \citealt{Tauris+2015} prescription for electron-capture SN) show that the primary NS tends to be more massive in the \Hecore channel.
}
\label{fig:masses_2channels}
\end{figure*}

\subsection{On ultra-stripped supernovae} \label{subsec:USSN}

Prior to the second SN, having undergone significant stripping, the progenitor is usually reduced to a metal-rich core with a thin envelope of mostly helium on top \citep{Yoon+2010}. The left panel in Figure \ref{fig:SN_progenitor} illustrates the mass distribution of this He layer for the two evolutionary sub-channels. Assuming this envelope mass can be used to represent the ejecta mass during core-collapse, we find that around $50\%$ of all binaries in the \Hecore channel have ejecta masses $M_{\rm ej} < 0.2 \Msun$, a commonly used criterion for an USSN \citep{Tauris+2015, De+2018}. The higher fraction of USSNs in the \Hecore subchannel can be attributed to the close proximity between the helium star and the NS (see Figure \ref{fig:CO-HeMS}), which grants the former less time to evolve before filling its Roche lobe and initiating Case BB mass transfer. 
As it is comparatively less-evolved, the helium star progenitor to the second NS undergoes RLO for more time prior to the termination of nuclear burning, resulting in a less massive envelope.
We find that most of our models produce a median helium envelope mass $\sim 0.25~\Msun$ at the time of the second SN, very similar to the value of $0.24~\Msun$ found by \cite{Jiang+2021} in their model of a typical USSN progenitor for a DNS system. 

Additionally, we note that around $60\%$ of the helium-rich progenitors were overfilling their Roche lobes during the time of the second SN explosion, as shown in the rightmost panel of Figure \ref{fig:SN_progenitor}. This is in agreement with previous studies of individual systems like \citet{Willems&Kalogera2004}, who argue using pre-SN orbital constraints that the helium star progenitor to the second NS in the double pulsar system was likely in a state of RLO at the time of the SN. 

For DNSs that merge within a Hubble time (formed through the \Hecore channel), approximately $\mathrm{96\%} \, (\mathrm{50\%})$ have pre-SN envelope masses less than $0.4\,\Msun \, (0.2\,\Msun)$, with none exceeding $1\,\Msun$. These values are significantly lower than the typical ejecta masses from core-collapse supernovae, which usually release a few solar masses of material. 
At the same time, Figure~\ref{fig:SN_progenitor} demonstrates that DNSs from the \COcore channel (which do not merge in a Hubble time) have larger envelope masses, making them unlikely to form from a USSN.
Thus, it is reasonable to expect the rate of USSN ought to closely follow the formation rate of galactic DNS systems \citep[][]{Tauris+2013} and be somewhat larger than the DNS merger rate. While some USSN could occur in NS--BH systems, we expect these to be a minority.

\subsection{On the observational classification}
\label{subsec:SN_types}

Since progenitors to both supernovae in our DNSs engage in mass transfer prior to collapse, we expect that the supernova forming them would be observed as Type Ib/c rather than Type II; indeed previous studies have suggested that binary interactions are the primary mechanism for producing Type Ib/c SN \citep[][]{Yoon+2010, Elridge+2013, Tauris+2013, Solar+2024}. Studies of synthetic spectra by \citet{Hachinger+2012} indicate that a minimum of $\mathrm{0.06~\Msun}$ of helium is necessary for helium lines to be observable in optical/IR spectra. We therefore assign collapsing stars with He envelopes less than $\mathrm{0.06~\Msun}$ as forming in a Type Ic SN. For systems with slightly more massive He-envelopes, their characteristics allow for some uncertainty in their observational classification \citep{dessart+2011}. We consider any star with a He-envelope mass greater than 0.1~$\Msun$ as a Type Ib SN. Between the two limits 0.06\,$\Msun$\,--\,0.1\,$\Msun$, we leave the classification as uncertain. 

Comparison with the He-envelope masses in the left panel of Figure~\ref{fig:SN_progenitor} shows that all the second born NSs formed through the \COcore channel will be formed within a Type Ib SN. However, for the \Hecore channel $\sim 90\%$ of the secondary SN will form within a Type Ib SN, 5\% as Type Ic, while the intermediate 5\% are uncertain. 

\subsection{On double neutron star masses} \label{sec:DNS_mass}

\begin{figure*}[ht!]
    \centering
    \includegraphics[width=1.0\linewidth]{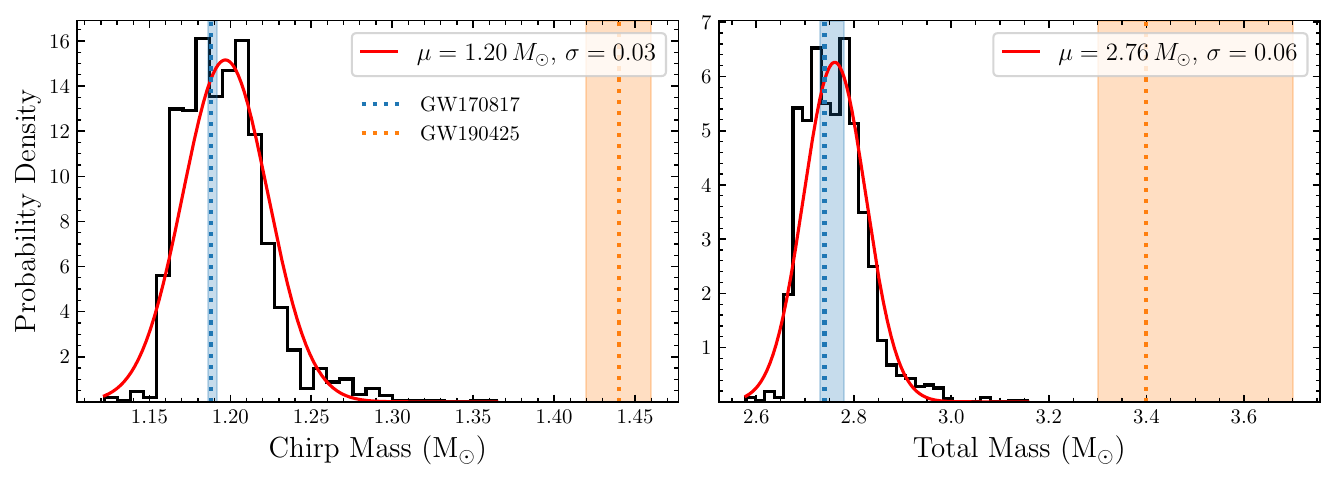}
    \caption{Distributions of the chirp mass \(\mathcal{M} = (m_1 m_2)^{3/5} (m_1 + m_2)^{-1/5}\) and total mass \(m_{\rm tot} = m_1 + m_2\) for merging DNSs from \texttt{MODEL12} (using the \citealt{Sukhbold+2016} prescription for core-collapse and \citealt{Tauris+2015} prescription for electron-capture SN) are shown along with Gaussian fits (red line) and the best-fit values. The source-frame chirp mass and total mass of the gravitational wave events GW170817 \citep{LIGOGW170817+2017} and GW190425 \citep{LIGOGW190425+2020} are plotted for reference.}
    \label{fig:chirp_and_total_mass}
\end{figure*}

\cite{Podsiadlowski+2004} and \cite{Heuvel2004} proposed that ECSN could lead to a population of NSs that are less massive than those from an Fe core collapse origin, and having a characteristic mass centered around $\simeq$ 1.25\Msun $\,$ \citep[see also][]{Schwab+2010}. For systems falling in the suitable C/O mass range for ECSN \citep{Tauris+2015}, we set the remnant baryonic mass of the proto-NS to be 1.38\Msun, which when converted into a  gravitational mass \,(Eq.\,\ref{eq:rembar}), results in a 1.26\Msun \, NS.
As discussed in Section \ref{subsec:natal_kicks}, our model populations strongly suggest that the first NS is likely to originate from an Fe core collapse (and therefore, should be the heavier one), while the second NS forms through an ECSN or a CCSN of an ultra-stripped core. This mass trend is illustrated in Figure \ref{fig:masses_2channels}, where we show the mass distribution of the recycled (left panel) and young NS (right panel) for \texttt{MODEL12}. The recycled pulsar has a mass in the range 1.25 -- 1.6$\Msun$, while the young pulsar is typically 1.26~\Msun, but can be as high as $\simeq1.5~\Msun$.
Additionally, the majority of the observed DNSs (Table \ref{tab:DNS_observations}) suggest that, in most cases where mass measurements exist, the first-born (recycled) NS in the binary is the heavier one (for example, the young, unrecycled NS in J1906$+$0746 is the observed pulsar, which appears to have a mass favoring an ECSN origin).

Detailed analysis of the observed DNSs has produced characteristic descriptions of the systems' masses \citep{Ozel+2016, Farrow+2019, Andrews2020}. The masses of our modeled systems do not match any of these descriptions particularly well due to limitations in our ability to calculate the resulting NS mass within our model; in \posydon we stop the evolution of stars upon core carbon exhaustion and adopt one of the available SN prescriptions to calculate the resulting NS mass. One should therefore be wary of over-interpreting the distributions in Figure~\ref{fig:masses_2channels}, instead using these as a qualitative guide for our model's results. With this caveat in mind, we see that among the first-born NSs forming from a CCSN, the \Hecore channel produces somewhat more massive NSs than the \COcore channel, as expected from the initially more massive progenitors of this channel. In principle this is a testable prediction: if more sub-population (ii) systems have their post-Keplerian parameters measured, allowing for precise component mass constraints, our model predicts that the recycled pulsars in these systems will be systematically less massive than those in sub-population (i).

So far, our analysis has focused on comparison with Galactic DNS systems. However, gravitational wave observatories have detected the merger of two DNSs, each of which have mass constraints \citep{LIGOGW170817+2017, LIGOGW190425+2020}. As extragalactic sources with entirely separate observational biases, these two events provide an alternative comparison to our models. GW170817 \citep{LIGOGW170817+2017}, with a source-frame chirp mass of $1.188^{+0.004}_{-0.002}\Msun$ and a total mass of $2.74^{+0.04}_{-0.01}\Msun$ falls within the 1-$\sigma$ confidence interval of our model predictions for merging DNSs (see Figure \ref{fig:chirp_and_total_mass}). However, none of our models predict DNSs with masses as high as that observed in GW190425 \citep[total mass $3.4^{+0.3}_{-0.1} \Msun$;][]{LIGOGW190425+2020}. Comparison with the Galactic DNSs in Table~\ref{tab:DNS_observations} show that, where they have been measured, none of the NSs are as massive as those that generated GW190425.

Several solutions have been proposed to explain the paucity of massive DNSs in the Milky Way sample, which generally fall into two camps. First, perhaps massive DNSs are formed at such short orbital periods that they quickly merge after formation, giving them a very short window for detection \citep{Safarzadeh+2019, IRS+2020, Galaudage+2021}. Such an ``ultra-fast merging'' channel likely requires systems to survive unstable Case BB mass transfer to produce sufficiently tight pre-SN orbits, a possibility precluded by our results in Section~\ref{subsec:CO-HeMS}. Alternatively, if the NS's natal kick is optimally directed, it could shrink the orbit, leading to a very short merger time \citep{Beniamini+2024}. Our results in Section~\ref{subsec:natal_kicks} suggest this is also an unlikely source of ``ultra-fast merging'' systems, as such kicks need to be of similar magnitude to the pre-SN orbital velocity, much larger than the 50-100 km/s kicks of our best models.

A second explanation for the lack of massive DNSs in the Milky Way sample proposes that massive NSs are formed with weak magnetic fields rendering such massive systems undetectable to our radio telescopes \citep{Safarzadeh+2020}. This could be due to SN fallback accretion that buries a newborn NS's magnetic field \citep{Alejandro+2021} or stable Case BB mass transfer and super-Eddington accretion leading to magnetic field burial \citep{Zhang+2023, Qin+2024, Chu+2025}. As our models do not include the fallback described in \citet{Alejandro+2021} nor do they include a pulsar model, we cannot currently test these scenarios. Given the significant uncertainties in the prescriptions that govern NS masses following stellar collapse, we here reiterate the need for caution while interpreting predictions about DNS masses.

\section{Merger Rates} 
\label{sec:merger_rates}

Following the methodology described in Section \ref{subsec:POSYDON}, we provide the local ($z=0$) merger rate for each of our tested models in the last column of Table~\ref{tab:rates}. The fraction of merging DNSs in any model is set by a combination of both the CE model, the ejection efficiency $\alpha_{\rm CE}$, as well as the natal kick prescription. We find that, with all other parameters held constant, the merger rate increases as we increase $\alpha_{\rm CE}$ or equivalently, increase the H-fraction that goes into defining the core-envelope boundary. This rate increases because in both cases, the \Hecore channel, comprising the merging population of DNSs, becomes more populated as the CE ejection process becomes more efficient. 

Additionally, when other factors are fixed, ``physics-based kick prescriptions" result in significantly higher merger rates compared to models adopting Maxwellian kicks from \citet{Hobbs+2005}. All of the Hobbs-Maxwellian kick models we test produce merger rates $\mathcal{O}(1)$ Gpc$^{-3}$ yr$^{-1}$, far smaller than either the DNS merger rate derived from observations of the sample of Galactic DNSs \citep[$293^{+222}_{-103} \rm{\,Gpc^{-3}\,yr^{-1}}$;][]{Kalogera+2004, O'Shaughnessy+Kim2010, Kim+2015, Pol+2019, Pol+2020, Grunthal+2021, Bernadich+2023} or the most recent DNS merger rate estimates following the first three observing runs of LIGO-Virgo-KAGRA Collaboration \citep[$10-1700\rm{\,Gpc^{-3}\,yr^{-1}}$;][]{GWTC-3}. These lower rates are caused by the Hobbs-Maxwellian kick models disrupting most systems during SN1. In contrast, most of our model populations with the ``physics-based kick prescriptions,'' produce merger rates in the range $\mathcal{O}(10)-\mathcal{O}(100)$ Gpc$^{-3}$ yr$^{-1}$, consistent with observational constraints from both gravitational wave observatories as well as the Milky Way DNS population.

Finally, we point out that previous binary population synthesis studies produce DNS merger rates that span several orders of magnitude, depending on the code and the adopted combination of binary parameters included \citep[e.g.,][]{ deMink&Belczynski2015, Chruslinska+2018, Alejandro+2018, Kruckow+2018}. Although a complete discussion describing the differences in each of these codes' merger rates is outside the scope of this work \citep[for a more comprehensive review of DNS merger rates, we refer to ][]{Mandel&Broekgaarden2022}, we discuss major differences between POSYDON and other codes in the following section.

\section{Comparison to previous binary population synthesis studies}
\label{sec:comparison}

Since our approach to binary population synthesis relies on self-consistent modeling of both stars' structures with the orbit, it is perhaps unsurprising that our results differ from previous population synthesis studies. While a comprehensive comparison to each of the previously published works is outside the scope of this work, we nevertheless provide a brief comparison to several previous studies of DNS formation.

The most important differences lie in the broadly defined DNS formation channels. While we find that $\gtrsim$ 95\% of all systems in \posydon form through the two subchannels shown in Figure~\ref{fig:schematic}, previous studies have identified additional scenarios. For instance, \citet{Andrews+2015} find that DNSs can additionally form through a CE without a subsequent stable MT phase. While \posydon does form systems though this channel (see the slight overlap in Figure~\ref{fig:CO-HeMS} between the \COcore channel outline and the gray region avoiding case BB mass transfer), they provide a very small contribution ($\lesssim5$\%) to the overall population (see Appendix~\ref{sec:app_form_chan}). As our systems tend to lie on the edge of the stable Case BB mass transfer regime, slight differences about the assumptions of helium star radii could be the source of these differences. 

In their study of DNS merger rates, \citet{Dominik+2013} make a distinction between an ``optimistic'' and ``pessimistic'' formation scenario, depending on whether Hertzsprung Gap (HG) stars are allowed to survive a CE. Since HG stars are still building their cores, it is unclear whether such stars would have a sufficiently steep density gradient to halt the inspiral of a companion \citep{deloye+2010}. As noted by \citet{Safarzadeh+2019}, the ``optimistic'' models from \citet{Dominik+2013} not only allow survival of H-rich HG stars through a CE, but also He-rich HG stars, the former being RLO from a traditional HG star while the latter corresponds to unstable Case BB MT. \posydon more closely matches the ``pessimistic'' scenario, as systems without a well-defined helium core merge within a CE\footnote{We note that our definitions of stellar types may differ; stars in \posydon are labeled based on their envelope composition and principal nuclear burning species rather than more traditional labels such as main sequence or Hertzsprung gap. However, the large radii of stars entering a CE, as shown in the bottom right panel of Figure~\ref{fig:oRLO2}, indicates these donors are giant stars.}, and
our results indicate that DNSs cannot form through unstable Case BB mass transfer (see Section~\ref{subsec:CO-HeMS}). 

Two other formation scenarios have been presented in the literature. First, \citet{Alejandro+2018} find that $\simeq$ 20\% of DNSs in their model form through a double-core CE. Our models find that this channel contributes at most $\simeq$ 2\% to the overall population of DNSs (see discussion in Section~\ref{sec:app_form_chan}). While the relative contribution from different formation channels has a highly non-linear dependence on binary evolution parameters, this scenario is especially dependent on the adopted radii of giant stars, which the fitting formula from \citet{hurley+2000} are known to significantly overestimate \citep{linden+2010, romagnolo+2023}. 
Second, \citet{Stevenson+2022} discuss the formation of very wide systems ($P_{\rm orb}\sim 10^3-10^5$ days) that avoid any mass transfer. We describe this scenario further in Appendix~\ref{sec:app_form_chan}, finding that \posydon indeed produces systems through this channel in agreement with \citet{Stevenson+2022}, but at a very small rate ($\lesssim$ 5\% of the overall population of DNSs). 

While these are just a few of the many binary population synthesis studies focusing on DNS formation in literature, we are not aware of any previous study that identifies a bifurcation in the resulting DNS population due to the evolutionary state of the donor upon RLO. In Section~\ref{sec:ce}, we attribute this bifurcation to the vast differences in the envelope binding energies between giant stars before helium ignition and after a C/O-core has formed. In principle, rapid binary population synthesis codes that either incorporate prescriptions for the envelope binding energies \citep[e.g., using the fitting formula from][]{xu+2010} or use grids of stellar models rather than fitting formula \citep[e.g., {\tt SEVN}, {\tt COMBINE}, or {\tt METISSE};][]{spera+2017, Kruckow+2018, agrawal+2020}, may be capable of capturing this bifurcation.

In addition to the formation channels, other groups have explored how assumptions about the CE influence the resulting DNS population. By comparing their models to some combination of the Galactic population and merger rates, these works generally find that the observations are best fit when $\alpha_{\rm CE}$ is larger than 1 \citep{Mapelli&Giacobbo2018, Chu+2022, Cecilia+2023, Deng+2024}. While we find that $\alpha_{\rm CE}=1$ can reproduce the observations, we require a generous definition of the CE core boundary, suggesting that a more flexible prescription rather than a large $\alpha_{\rm CE}$ value alone may be consistent with their findings. 

Finally, previous studies investigating SN kicks tend to find that smaller kicks better match the observed distribution of DNSs in the Milky Way \citep{Andrews+2015, Kruckow+2018, Giacobbo&Mapelli2018, Shao&Li2018}. For instance, \citet{Alejandro+2018} found that a double-peaked distribution with a low-kick peak around $\sigma\simeq30$ km/s best fits the observations \citep[see also][]{Chu+2022}. Our conclusion that low-velocity kicks are required, as discussed in Section~\ref{subsec:natal_kicks}, is broadly consistent with these previous findings.

\section{Conclusions} \label{sec:conclusions}
We have placed constraints on the key evolutionary stages leading up to DNS formation using \posydon as our population synthesis tool and comparing our models with the observations of 25 DNS systems in the Milky Way field. 
While a thorough Bayesian analysis would be ideal for comparison with the observations, we set aside this approach due to the current absence of a pulsar evolution model within \posydon, opting instead for qualitative comparison of our model distributions with the observed DNSs in the Milky Way. We have run 25 separate \posydon models, exploring variations in the prescriptions describing common envelope evolution and the core-collapse which we find to have the strongest effects on our resulting populations. By comparing these evolutionary stages in detail, we are able to draw several conclusions about the formation of DNSs:

\begin{itemize}

\item Although other rare scenarios exist, our model populations show that almost all ($\gtrsim$ 95\%) DNS formation proceeds via a CE event, but bifurcated into one of two subchannels depending on the evolutionary state of the donor at the onset of CE with the first born NS.

\item One subchannel is characterized by a He core with a strong envelope binding energy, which requires an efficient CE to form: either the system detaches with a generous core definition of $\simeq$ 30\% or the CE evolves with a high CE ejection efficiency $\alpha_{\rm CE} \gtrsim 1.2$. In the other channel, donor stars fill their Roche lobes with a carbon-oxygen core and a loosely bound envelope, allowing survival even with low values of $\alpha_{\rm CE}$.

\item DNSs from these two subchannels differ in their distribution in the $P_{\rm orb}-e$ diagram, the recycling efficiency of the first born pulsar, as well as the mass of the primary NS. 
Following the division proposed by \citet{Andrews&Mandel2019} for the observed Milky Way DNSs, our model is suggestive that sub-population (i) systems formed through the \Hecore subchannel while sub-population (ii) systems formed through the \COcore subchannel.

\item Sub-population (iii) systems, while falling within the 3-$\sigma$ confidence interval for our models, remains difficult to explain due to the observed eccentricity gap between 0.3 $< e <$ 0.6. This population poses a challenge for isolated binary evolution models, suggesting that an alternative evolutionary pathway such as dynamical formation, tidal effects, magnetic field interactions, or circumbinary disk evolution, may be needed to account for its characteristics \citep[see][]{Andrews&Mandel2019}.

\item DNSs formed through the \Hecore channel tend to merge within a Hubble time while DNSs formed through the \COcore channel do not. Therefore, our models suggest that gravitational wave sources such as GW170817 and GW190425 are formed through the \Hecore channel. 

\item Following CE ejection, mass transfer with a He-star donor (so-called Case BB mass transfer) is necessary to produce merging DNSs. In agreement with previous studies \citep[e.g.,][]{Ivanova+2003, Tauris+2015}, this phase must be stable to lead to DNS formation; our detailed binary evolution grids find that if the binary enters unstable Case BB evolution, it either merges within the CE or creates a NS--WD system.

\item The primary SN in the system is driven by iron core-collapse, whereas the secondary SN, being subjected to binary interactions, has more NSs forming through an ECSN. The first-born NSs in DNSs tend to be systematically more massive than their companions, in broad agreement with the observed sample.

\item To simultaneously match the $P_{\rm orb}-e$ distribution of Galactic DNSs as well as the merger rate, we find that a Maxwellian kick distribution with a $\sigma=265$ km s$^{-1}$ is ruled out in agreement with previous results \citep[e.g.,][]{Bray&Eldridge2016}. This prescription disrupts most binaries during the first core collapse. Instead, we prefer a kick prescription with lower kick velocities; our default \textit{AsymEj} kick model, motivated by results from \citet{Janka2017}, produces kicks of 10-90 km/s for the first SN and $<$40 km/s for the second SN.

\item For approximately 60\% of our DNS systems, the second-born NS was overfilling its Roche lobe as a He giant when it went through core collapse, a phenomenon described by previous studies of individual DNSs \citep[e.g.,][]{Willems&Kalogera2004}.

\item The second-born NSs in the subset of systems that will merge in a Hubble time have lost most of their envelope mass prior to collapse. We find that up to 10\% of such systems could be observed as a Type Ic SN. 
Additionally, merging DNSs are typically born with a pre-SN He envelope mass $<$ 0.4 \Msun.

\item The volumetric DNS merger rate in the local universe assuming the \texttt{IllustrisTNG} star formation history is estimated to be $\mathcal{O}(10)-\mathcal{O}(100)$ Gpc$^{-3}$ yr$^{-1}$, consistent with recent estimates from gravitational wave observatories \citep{GWTC-3} as well as limits from the sample of Galactic DNSs \citep{Bernadich+2023}. However, the merger rate predicted by \posydon varies substantially depending on the chosen model.

\end{itemize}

Despite the conclusions we draw from our models, one should be aware of a number of caveats associated with our results. First, \posydon is intrinsically limited by the density of our binary grids and the adopted interpolation scheme. We assume that accumulated errors are smaller than those associated with the fitting formula methods used by traditional BPS codes, although a detailed comparison between our code and others is outside the scope of this work. Furthermore, the models described in this study are limited to Solar metallicity. While most of the observed Galactic DNSs are young (see Table~\ref{tab:DNS_observations}) and therefore likely formed with compositions close to the Solar value, a more complete exploration will account for the contribution to the observed population due to lower Z systems. There is some evidence that different formation scenarios may become relevant at lower metallicities \citep[e.g.,][]{Chruslinska+2018, 2023NatAs...7..444S}. In future work we plan to explore DNS formation across a cosmological distribution of metallicities and star formation rates using \posydon v2 \citep[][]{Andrews+2025}. 

Finally, our study does not include a pulsar evolution model. An accurate modeling of the pulsar requires a detailed treatment of its magnetic field and spin evolution through the detached and recycling phases, accounting for selection effects \citep{Chattopadhyay+2020}. The inclusion of such a model, which we plan to develop in future work, opens the door for a more quantitative, Bayesian comparison with the Milky Way DNS sample. Nevertheless, our current work provides a foundation upon which future \posydon studies of DNS formation can be built.

\acknowledgments

We thank the anonymous referee for a careful review of the manuscript and providing valuable feedback that helped improve the quality of this paper. The \posydon{} project is supported primarily by two sources: the Gordon and Betty Moore Foundation (PI Kalogera, grant awards GBMF8477 and GBMF12341) and a Swiss National Science Foundation (PI Fragos, project numbers PP00P2\_211006 and CRSII5\_213497). 
AC gratefully acknowledges support from the Steigleman Family Fellowship, awarded by the University of Florida. 
JJA acknowledges support for Program number (JWST-AR-04369.001-A) provided through a grant from the STScI under NASA contract NAS5-03127. 
DC, PMS, KAR, and MS is supported by the project numbers GBMF8477 and GBMF12341.
KAR is also supported by the NASA grant awarded to the Illinois/NASA Space Grant Consortium, and any opinions, findings, conclusions, or recommendations expressed in this material are those of the author and do not necessarily reflect the views of NASA.
KK and EZ were partially supported by the Federal Commission for Scholarships for Foreign Students for the Swiss Government Excellence Scholarship (ESKAS No.~2021.0277 and ESKAS No.~2019.0091, respectively).
KK is supported by a fellowship program at the Institute of Space Sciences (ICE-CSIC) funded by the program Unidad de Excelencia Mar\'ia de Maeztu CEX2020-001058-M.
EZ acknowledges support from the Hellenic Foundation for Research and Innovation (H.F.R.I.) under the “3rd Call for H.F.R.I. Research Projects to support Post-Doctoral Researchers” (Project No: 7933). 
KAR thanks the LSSTC Data Science Fellowship Program, which is funded by LSSTCorporation, NSF Cybertraining Grant No. 1829740, the Brinson Foundation, and the Gordon and Betty Moore Foundation; their participation in the program has benefited this work.
ZX was supported by the Chinese Scholarship Council (CSC). 
The authors acknowledge UFIT Research Computing \url{http://www.rc.ufl.edu} for providing computational resources and support that have contributed to the research results reported in this publication. 
This research was supported in part through the computational resources and staff contributions provided for the Quest high performance computing facility at Northwestern University which is jointly supported by the Office of the Provost, the Office for Research, and Northwestern University Information Technology.

\software{This manuscript has made use of the following Python modules: 
\texttt{numpy} \citep{2020NumPy-Array},
\texttt{scipy} \citep{scipy},
\texttt{pandas} \citep{pandas},
\texttt{matplotlib} \citep{matplotlib},
\texttt{astropy} 
\citep{astropy_I, astropy_II, astropy_III}
\texttt{scikit-learn} \citep{scikit-learn}. 
}

\newpage

\appendix
\section{Classification of formation channels}
\label{sec:app_form_chan}

From our model populations in Table \ref{tab:rates}, we identify four evolutionary tracks leading to DNS formation. While we only discuss the first one (common envelope channel) in the main text as it dominates DNS formation, we describe three other channels below that contribute at a negligible level.

\begin{itemize}
    \item Common envelope channel: This is by far the most dominant among our formation channels, accounting for $\gtrsim 95\%$ of all DNS systems in our model populations. This channel is characterized by a CE phase of the first-born NS with the H-rich companion. Subsequently, the stripped core of the donor undergoes expansion into a helium giant, initiating stable Case BB RLO. Only a small fraction ($<5\%$) avoid mass transfer with the He star. 
    
    \item Non-interacting channel: A second formation channel, responsible for upto $\sim 5\%$ of all DNSs in our model populations, manage to entirely avoid any kind of mass transfer. Despite the two SN explosions, they are able to survive in very wide ($P_{\rm orb} \sim 10^3 - 10^5$ days) and highly eccentric orbits. \citet{Stevenson+2022} discuss a population of wide, non-recycled binary pulsars formed via ECSN mechanism, and although our models indicate such systems originate from Fe core-collapse,
    this \textit{non-interacting} channel could potentially represent a similar population. In this context, the ultra-wide pulsar reported in \cite{VanDerWateren+2024} could be a potential system from this population.
    
    \item Stable RLO channel: A third channel, although very rare ($<1\%$), involves only stable mass transfer from the H-rich donor, the He-rich donor, or both, onto the first born NS. DNSs born through this channel have orbital periods falling somewhere between the first two channels ($P_{\rm orb} \sim 1 - 10^3$ days). 

    \item Double-core common envelope channel: The fourth formation pathway proceeds through a double-core common envelope phase between the two initial ZAMS stars. This channel occurs for systems with similar initial masses such that when one star overfills its Roche lobe, the other already has a large, high-entropy envelope, having evolved off the main sequence. This channel contributes at most $\sim2\%$ of DNS systems in our model populations. 
\end{itemize}

\section{Justifying the exclusion of wind accretion}
\label{sec:appendix_wind_accretion}

\begin{figure}
    \centering
    \includegraphics[width=1.0\linewidth]{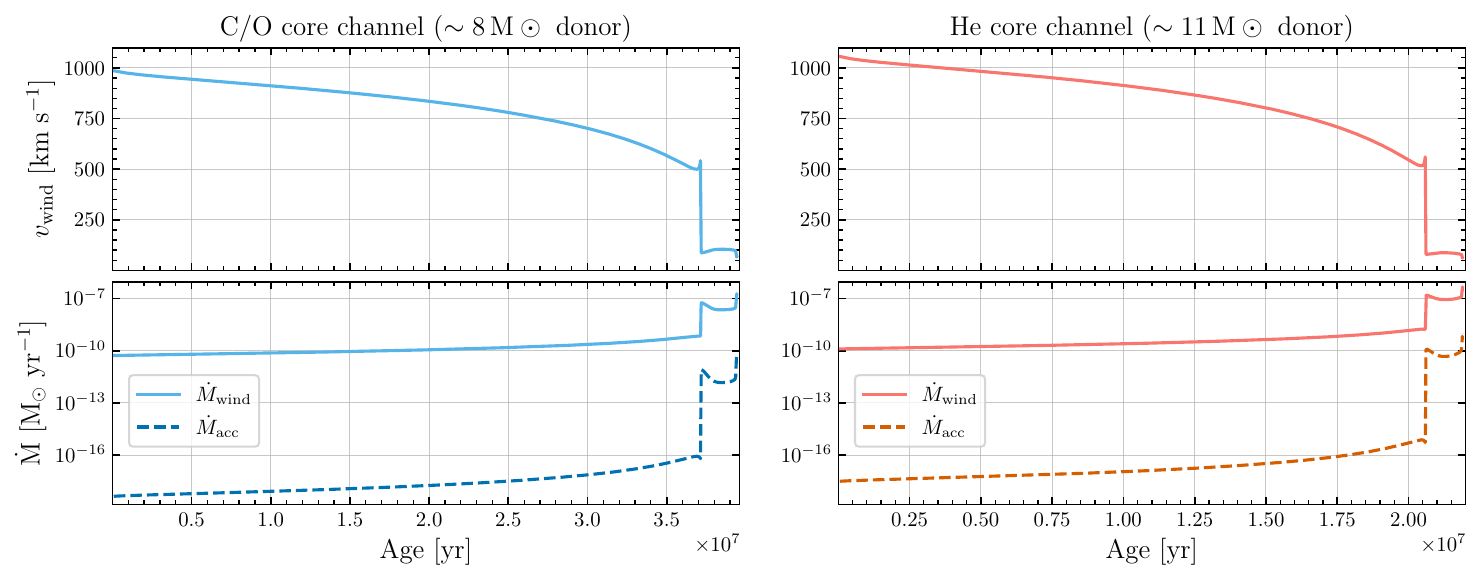}
    \caption{Evolution of the terminal wind velocity (top panels) and the corresponding Bondi-Hoyle-Lyttleton accretion rate onto a 1.4\Msun\,NS (bottom panels) calculated for the \COcore channel (left) and \Hecore channel (right). Typical donor masses and binary separations in the two channels have been assumed.}
    \label{fig:wind_mass_loss}
\end{figure}

To estimate the potential contribution of wind accretion onto the first-born NS, we consider spherically symmetric winds with speed $v_{\rm wind}$ from the companion. The Bondi-Hoyle-Lyttleton \citep[BHL;][]{Hoyle&Lyttleton1939, Bondi&Hoyle1944, Bondi1952} accretion rate is given by:
\begin{equation}
\begin{split}
\label{eq:mdot_bhl}
\dot{M}_{{\rm acc}}	&=\frac{G^{2}M_{{\rm NS}}^{2}\dot{M}_{{\rm wind}}}{v_{{\rm wind}}^{4} a^{2}} \\
\end{split}
\end{equation}
where $\dot{M}_{\rm wind}$ is the mass-loss rate from the donor, which is approximated using the prescription from \citet{Reimers1975}:
\begin{equation}
    \label{eq:mdot_reimers}
    \dot{M}_{{\rm wind}}=4\times 10^{-13}\beta_{R}M_{\odot}\,{\rm yr}^{-1}\left(\frac{L_\star}{L_{\odot}}\right)\left(\frac{R_\star}{R_{\odot}}\right)\left(\frac{M_\star}{M_{\odot}}\right)^{-1}
\end{equation}
where $\beta_R$ is a dimensionless constant, and we adopt $\beta_R=0.1$ following \citet{Choi+2016}. The wind velocity is then calculated at each timestep in the star's evolution, assuming it to be equal to the escape velocity from the stellar surface \( v = \sqrt{2 G M_\star / R_\star} \). Figure \ref{fig:wind_mass_loss} displays the resulting terminal wind velocities and corresponding BHL accretion rates from a single H-rich star onto a 1.4\Msun \ NS, assuming representative orbital separations and donor masses for the \Hecore and \COcore channels. 

As an upper limit on the amount of mass accreted via winds, we integrate over the wind accretion curve $\dot M_{\rm acc}$ in Figure \ref{fig:wind_mass_loss} under the assumption of fully conservative accretion, finding that the total wind mass accreted by the BHL mechanism is $\Delta M_{\rm acc} \simeq 10^{-5}\,\Msun$ in the \COcore channel, and $\Delta M_{\rm acc} \simeq 6 \times 10^{-5}\,\Msun$ for the \Hecore channel. We are thus led to conclude that contribution from wind accretion can be neglected compared to Roche lobe overflow, which typically yields accreted masses of $\sim 10^{-4} - 10^{-3} \Msun$ (see Figure \ref{fig:mass_accreted}). 

\section{The Initial Conditions for DNS Formation}
\label{sec:HMS_HMS}

\begin{figure*}[ht!]
\includegraphics[width=1.0\linewidth]{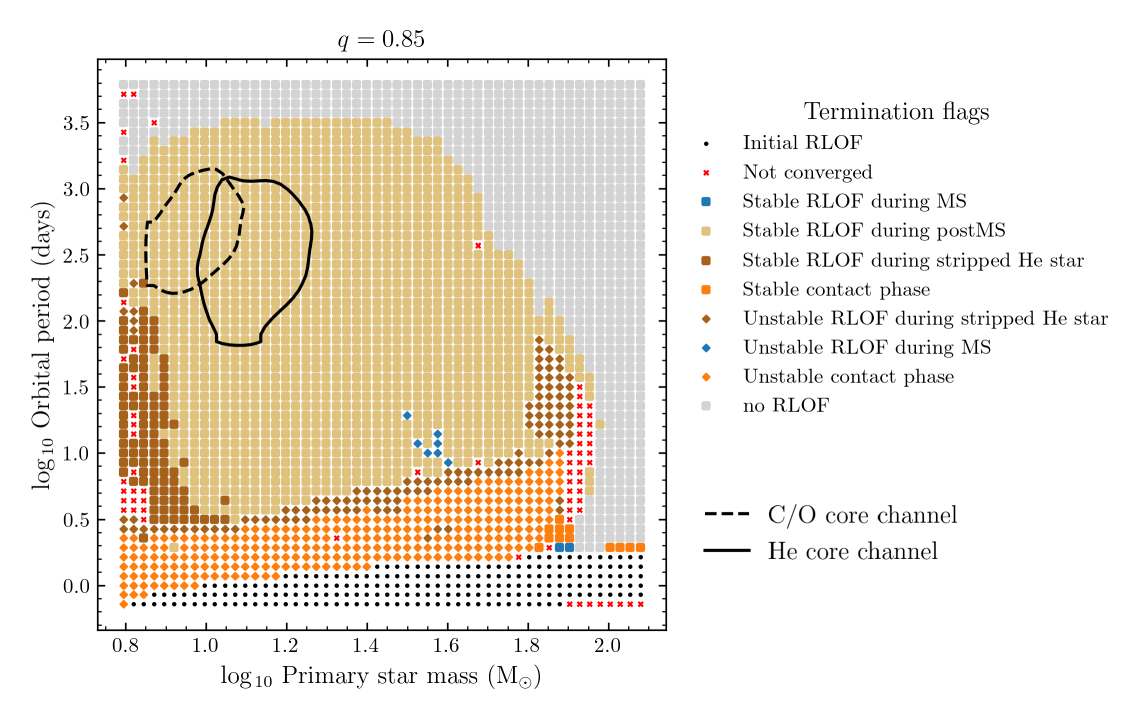}
\caption{A 2D slice with different markers summarizing the evolution of systems in our \texttt{HMS-HMS} grid comprising of two hydrogen-rich stars initially at ZAMS. The two black contours enclose $75\%$ of the primary ZAMS stars that form DNSs through the two sub-channels in \texttt{MODEL12}. The slice has been generated at $q =$ 0.85 to reflect the mean mass ratio at this stage. Notably, the initial binaries in the \Hecore channel tend to have more massive primaries and shorter orbital periods compared to those in the \COcore channel.
\label{fig:HMS-HMS}}
\end{figure*}

We described in Sections \ref{subsec:CO-HMS_RLO} and \ref{subsec:CO-HeMS} the detailed evolutionary pathways of DNS-forming binaries through the NS+H-rich star and NS+He-rich star phases, respectively. Here, we include a corresponding description to illustrate how these systems originate from binaries composed of two H-rich stars at ZAMS. Figure \ref{fig:HMS-HMS} displays a two-dimensional slice of our \texttt{HMS-HMS} grid. Each marker represents a separate simulation with varying primary mass and orbital period, while maintaining a fixed initial mass ratio of $q = 0.85$. This value was chosen to represent the typical mass ratio between the initial ZAMS stars in our models that lead to DNS formation, with the majority having $q \gtrsim 0.7$.

The dashed and solid contours enclose 75\% of the initial systems from the two sub-channels. The initial primary is an OB star, 
typically more massive in the \Hecore channel (median $\sim$ 12.9\Msun) than in the \COcore channel (median $\sim$ 9.6\Msun). The \Hecore systems originate from binaries with $P_{\rm orb} \sim$ 300 days, while the \COcore channel typically has $P_{\rm orb} \sim$ 500 days. 
The majority of these systems undergo stable mass transfer (tan square markers) during the post H-burning expansion phase of the primary.
Notably, no DNSs are produced from binaries that underwent a CE phase between their H-rich components (diamond markers). We find that the maximum primary mass at ZAMS leading to DNS formation is approximately $25\Msun$, with a companion of $\sim 17\Msun$. 
\label{sec:appendix}

\bibliographystyle{aasjournal}
\bibliography{references}

\end{document}